\documentclass[11pt]{article}

\usepackage{xcolor}
\usepackage{titlesec}
\usepackage{abstract}
\usepackage{multicol}
\usepackage{tocloft}
\usepackage{graphicx}
\usepackage[footnotesize,hang,bf]{caption}
\usepackage[font=footnotesize]{subcaption}
\usepackage{times}
\usepackage{amssymb,amsmath}
\usepackage{footmisc}
\usepackage{appendix}
\usepackage{epstopdf}
\usepackage[leftcaption]{sidecap}
\usepackage{wrapfig}
\usepackage{enumerate}
\usepackage{fancyhdr}
\usepackage{makeidx}
\usepackage{indentfirst}
\usepackage{multirow}
\usepackage{color}
\usepackage{fancybox}
\usepackage{listing}
\usepackage[hyphens]{url}
\usepackage[intoc]{nomencl}
\usepackage{array}
\usepackage{cfxminted}

\newcommand{\tighterenum}{\setlength{\parskip}{-1pt}}

\definecolor{mtcode}{rgb}{1.0, 1.0, 1.0}

\titleformat{\section}{\raggedright\bfseries\large\scshape}{\thesection.}{0.5em}{}
\titleformat{\subsection}{\raggedright\normalfont\scshape}{\thesubsection.}{0.5em}{}
\titleformat{\subsubsection}{\raggedright\normalfont\small\scshape}{\thesubsubsection.}{0.5em}{}

\definecolor{linkcolor}{rgb}{0.00,0.00,0.75}
\definecolor{citecolor}{rgb}{0.00,0.00,0.75}
\definecolor{linkbordercolor}{rgb}{0.00,0.00,0.75}

\usepackage{hyperref}
\hypersetup{%
  linktocpage=true,
  colorlinks=true,
  urlcolor=linkcolor, 
  linkcolor=linkcolor,
  menucolor=linkcolor,
  bookmarks=true,    
  bookmarksopen=true,    
  hyperfootnotes=false,    
  pdfpagemode=UseOutlines,    
  citecolor=citecolor,
  linkbordercolor=linkbordercolor,
  citebordercolor=linkbordercolor,
  urlbordercolor=linkbordercolor,
      pdftitle={Intel Cilk Plus for Complex Parallel Algorithms: "Enormous Fast Fourier Transforms" (EFFT) Library},
      pdfsubject={Paper},
      pdfauthor={Ryo Asai and Andrey Vladimirov},
      pdfkeywords={Cilk Plus, MKL, FFTW, EFFT, bandwidth, xeon, library, optimization}
}
\makeatletter
\Hy@AtBeginDocument{%
  \def\@pdfborder{0 0 1}
  \def\@pdfborderstyle{/S/U/W 0.6}
}
\makeatother

\makeatletter
\def\PY@reset{\let\PY@it=\relax \let\PY@bf=\relax%
    \let\PY@ul=\relax \let\PY@tc=\relax%
    \let\PY@bc=\relax \let\PY@ff=\relax}
\def\PY@tok#1{\csname PY@tok@#1\endcsname}
\def\PY@toks#1+{\ifx\relax#1\empty\else%
    \PY@tok{#1}\expandafter\PY@toks\fi}
\def\PY@do#1{\PY@bc{\PY@tc{\PY@ul{%
    \PY@it{\PY@bf{\PY@ff{#1}}}}}}}
\def\PY#1#2{\PY@reset\PY@toks#1+\relax+\PY@do{#2}}

\expandafter\def\csname PY@tok@gd\endcsname{\def\PY@tc##1{\textcolor[rgb]{0.63,0.00,0.00}{##1}}}
\expandafter\def\csname PY@tok@gu\endcsname{\let\PY@bf=\textbf\def\PY@tc##1{\textcolor[rgb]{0.50,0.00,0.50}{##1}}}
\expandafter\def\csname PY@tok@gt\endcsname{\def\PY@tc##1{\textcolor[rgb]{0.00,0.27,0.87}{##1}}}
\expandafter\def\csname PY@tok@gs\endcsname{\let\PY@bf=\textbf}
\expandafter\def\csname PY@tok@gr\endcsname{\def\PY@tc##1{\textcolor[rgb]{1.00,0.00,0.00}{##1}}}
\expandafter\def\csname PY@tok@cm\endcsname{\let\PY@it=\textit\def\PY@tc##1{\textcolor[rgb]{0.25,0.50,0.50}{##1}}}
\expandafter\def\csname PY@tok@vg\endcsname{\def\PY@tc##1{\textcolor[rgb]{0.10,0.09,0.49}{##1}}}
\expandafter\def\csname PY@tok@m\endcsname{\def\PY@tc##1{\textcolor[rgb]{0.40,0.40,0.40}{##1}}}
\expandafter\def\csname PY@tok@mh\endcsname{\def\PY@tc##1{\textcolor[rgb]{0.40,0.40,0.40}{##1}}}
\expandafter\def\csname PY@tok@go\endcsname{\def\PY@tc##1{\textcolor[rgb]{0.53,0.53,0.53}{##1}}}
\expandafter\def\csname PY@tok@ge\endcsname{\let\PY@it=\textit}
\expandafter\def\csname PY@tok@vc\endcsname{\def\PY@tc##1{\textcolor[rgb]{0.10,0.09,0.49}{##1}}}
\expandafter\def\csname PY@tok@il\endcsname{\def\PY@tc##1{\textcolor[rgb]{0.40,0.40,0.40}{##1}}}
\expandafter\def\csname PY@tok@cs\endcsname{\let\PY@it=\textit\def\PY@tc##1{\textcolor[rgb]{0.25,0.50,0.50}{##1}}}
\expandafter\def\csname PY@tok@cp\endcsname{\def\PY@tc##1{\textcolor[rgb]{0.74,0.48,0.00}{##1}}}
\expandafter\def\csname PY@tok@gi\endcsname{\def\PY@tc##1{\textcolor[rgb]{0.00,0.63,0.00}{##1}}}
\expandafter\def\csname PY@tok@gh\endcsname{\let\PY@bf=\textbf\def\PY@tc##1{\textcolor[rgb]{0.00,0.00,0.50}{##1}}}
\expandafter\def\csname PY@tok@ni\endcsname{\let\PY@bf=\textbf\def\PY@tc##1{\textcolor[rgb]{0.60,0.60,0.60}{##1}}}
\expandafter\def\csname PY@tok@nl\endcsname{\def\PY@tc##1{\textcolor[rgb]{0.63,0.63,0.00}{##1}}}
\expandafter\def\csname PY@tok@nn\endcsname{\let\PY@bf=\textbf\def\PY@tc##1{\textcolor[rgb]{0.00,0.00,1.00}{##1}}}
\expandafter\def\csname PY@tok@no\endcsname{\def\PY@tc##1{\textcolor[rgb]{0.53,0.00,0.00}{##1}}}
\expandafter\def\csname PY@tok@na\endcsname{\def\PY@tc##1{\textcolor[rgb]{0.49,0.56,0.16}{##1}}}
\expandafter\def\csname PY@tok@nb\endcsname{\def\PY@tc##1{\textcolor[rgb]{0.00,0.50,0.00}{##1}}}
\expandafter\def\csname PY@tok@nc\endcsname{\let\PY@bf=\textbf\def\PY@tc##1{\textcolor[rgb]{0.00,0.00,1.00}{##1}}}
\expandafter\def\csname PY@tok@nd\endcsname{\def\PY@tc##1{\textcolor[rgb]{0.67,0.13,1.00}{##1}}}
\expandafter\def\csname PY@tok@ne\endcsname{\let\PY@bf=\textbf\def\PY@tc##1{\textcolor[rgb]{0.82,0.25,0.23}{##1}}}
\expandafter\def\csname PY@tok@nf\endcsname{\def\PY@tc##1{\textcolor[rgb]{0.00,0.00,1.00}{##1}}}
\expandafter\def\csname PY@tok@si\endcsname{\let\PY@bf=\textbf\def\PY@tc##1{\textcolor[rgb]{0.73,0.40,0.53}{##1}}}
\expandafter\def\csname PY@tok@s2\endcsname{\def\PY@tc##1{\textcolor[rgb]{0.73,0.13,0.13}{##1}}}
\expandafter\def\csname PY@tok@vi\endcsname{\def\PY@tc##1{\textcolor[rgb]{0.10,0.09,0.49}{##1}}}
\expandafter\def\csname PY@tok@nt\endcsname{\let\PY@bf=\textbf\def\PY@tc##1{\textcolor[rgb]{0.00,0.50,0.00}{##1}}}
\expandafter\def\csname PY@tok@nv\endcsname{\def\PY@tc##1{\textcolor[rgb]{0.10,0.09,0.49}{##1}}}
\expandafter\def\csname PY@tok@s1\endcsname{\def\PY@tc##1{\textcolor[rgb]{0.73,0.13,0.13}{##1}}}
\expandafter\def\csname PY@tok@sh\endcsname{\def\PY@tc##1{\textcolor[rgb]{0.73,0.13,0.13}{##1}}}
\expandafter\def\csname PY@tok@sc\endcsname{\def\PY@tc##1{\textcolor[rgb]{0.73,0.13,0.13}{##1}}}
\expandafter\def\csname PY@tok@sx\endcsname{\def\PY@tc##1{\textcolor[rgb]{0.00,0.50,0.00}{##1}}}
\expandafter\def\csname PY@tok@bp\endcsname{\def\PY@tc##1{\textcolor[rgb]{0.00,0.50,0.00}{##1}}}
\expandafter\def\csname PY@tok@c1\endcsname{\let\PY@it=\textit\def\PY@tc##1{\textcolor[rgb]{0.25,0.50,0.50}{##1}}}
\expandafter\def\csname PY@tok@kc\endcsname{\let\PY@bf=\textbf\def\PY@tc##1{\textcolor[rgb]{0.00,0.50,0.00}{##1}}}
\expandafter\def\csname PY@tok@c\endcsname{\let\PY@it=\textit\def\PY@tc##1{\textcolor[rgb]{0.25,0.50,0.50}{##1}}}
\expandafter\def\csname PY@tok@mf\endcsname{\def\PY@tc##1{\textcolor[rgb]{0.40,0.40,0.40}{##1}}}
\expandafter\def\csname PY@tok@err\endcsname{\def\PY@bc##1{\setlength{\fboxsep}{0pt}\fcolorbox[rgb]{1.00,0.00,0.00}{1,1,1}{\strut ##1}}}
\expandafter\def\csname PY@tok@kd\endcsname{\let\PY@bf=\textbf\def\PY@tc##1{\textcolor[rgb]{0.00,0.50,0.00}{##1}}}
\expandafter\def\csname PY@tok@ss\endcsname{\def\PY@tc##1{\textcolor[rgb]{0.10,0.09,0.49}{##1}}}
\expandafter\def\csname PY@tok@sr\endcsname{\def\PY@tc##1{\textcolor[rgb]{0.73,0.40,0.53}{##1}}}
\expandafter\def\csname PY@tok@mo\endcsname{\def\PY@tc##1{\textcolor[rgb]{0.40,0.40,0.40}{##1}}}
\expandafter\def\csname PY@tok@kn\endcsname{\let\PY@bf=\textbf\def\PY@tc##1{\textcolor[rgb]{0.00,0.50,0.00}{##1}}}
\expandafter\def\csname PY@tok@mi\endcsname{\def\PY@tc##1{\textcolor[rgb]{0.40,0.40,0.40}{##1}}}
\expandafter\def\csname PY@tok@gp\endcsname{\let\PY@bf=\textbf\def\PY@tc##1{\textcolor[rgb]{0.00,0.00,0.50}{##1}}}
\expandafter\def\csname PY@tok@o\endcsname{\def\PY@tc##1{\textcolor[rgb]{0.40,0.40,0.40}{##1}}}
\expandafter\def\csname PY@tok@kr\endcsname{\let\PY@bf=\textbf\def\PY@tc##1{\textcolor[rgb]{0.00,0.50,0.00}{##1}}}
\expandafter\def\csname PY@tok@s\endcsname{\def\PY@tc##1{\textcolor[rgb]{0.73,0.13,0.13}{##1}}}
\expandafter\def\csname PY@tok@kp\endcsname{\def\PY@tc##1{\textcolor[rgb]{0.00,0.50,0.00}{##1}}}
\expandafter\def\csname PY@tok@w\endcsname{\def\PY@tc##1{\textcolor[rgb]{0.73,0.73,0.73}{##1}}}
\expandafter\def\csname PY@tok@kt\endcsname{\def\PY@tc##1{\textcolor[rgb]{0.69,0.00,0.25}{##1}}}
\expandafter\def\csname PY@tok@ow\endcsname{\let\PY@bf=\textbf\def\PY@tc##1{\textcolor[rgb]{0.67,0.13,1.00}{##1}}}
\expandafter\def\csname PY@tok@sb\endcsname{\def\PY@tc##1{\textcolor[rgb]{0.73,0.13,0.13}{##1}}}
\expandafter\def\csname PY@tok@k\endcsname{\let\PY@bf=\textbf\def\PY@tc##1{\textcolor[rgb]{0.00,0.50,0.00}{##1}}}
\expandafter\def\csname PY@tok@se\endcsname{\let\PY@bf=\textbf\def\PY@tc##1{\textcolor[rgb]{0.73,0.40,0.13}{##1}}}
\expandafter\def\csname PY@tok@sd\endcsname{\let\PY@it=\textit\def\PY@tc##1{\textcolor[rgb]{0.73,0.13,0.13}{##1}}}

\def\PYZus{\char`\_}
\def\PYZob{\char`\{}
\def\PYZcb{\char`\}}

\def\PYZam{\char`\&}
\def\PYZlt{\char`\<}
\def\PYZgt{\char`\>}
\def\PYZsh{\char`\#}

\def\PYZhy{\char`\-}

\def\PYZti{\char`\~}

\makeatother

\newcommand{\TM}{\textsuperscript{\texttrademark}}

\definecolor{quotecolor}{RGB}{255,255,200}

\ifdefined\mobiledoc
 \newcommand{\optcolstart}{}
 \newcommand{\optcolend}{}
 \newcommand{\optcolsep}{\newpage}
 \newcommand{\colfaxwidth}{2.8 in}
 \fancyfootoffset{-2pt}
\else
 \newcommand{\optcolstart}{\begin{multicols}{2}}
 \newcommand{\optcolend}{\end{multicols}}
 \newcommand{\optcolsep}{\vfill\columnbreak}
 \newcommand{\colfaxwidth}{5.7 in}
\fi

\makenomenclature

\begin{document}
\sloppy

{\onecolumn

\author{\slshape Ryo Asai and Andrey Vladimirov \\ \slshape Colfax International}

\title{\Large\sffamily\scshape
\vskip -2.5 em
Intel Cilk Plus for Complex Parallel Algorithms:\\
``Enormous Fast Fourier Transforms'' (EFFT) Library\\
}

\date{\small September 18, 2014}

{\linespread{0.8} \maketitle}

\optcolstart

\begin{abstract}
\normalsize

In this paper we demonstrate the methodology for parallelizing the computation of large one-dimensional discrete fast Fourier transforms (DFFTs) on multi-core Intel Xeon processors.
DFFTs based on the recursive Cooley-Tukey method have to control cache utilization, memory bandwidth and vector hardware usage, and at the same time scale across multiple threads or compute nodes.
Our method builds on single-threaded Intel Math Kernel Library (MKL) implementation of DFFT, and uses the Intel Cilk Plus framework for thread parallelism.
We demonstrate the ability of Intel Cilk Plus to handle parallel recursion with nested loop-centric parallelism without tuning the code to the number of cores or cache metrics.
The result of our work is a library called EFFT that performs 1D DFTs of size $2^{N}$ for $N\geq21$ faster than the corresponding Intel MKL parallel DFT implementation by up to 1.5x, and faster than FFTW by up to 2.5x. The code of EFFT is available for free download under the GPLv3 license.

This work provides a new efficient DFFT implementation, and at the same time demonstrates an educational example of how computer science problems with complex parallel patterns can be optimized for high performance using the Intel Cilk Plus framework.

\phantom{ \cite{efft} }

\end{abstract}

\optcolsep

{\small
\tableofcontents

}

\optcolend

\vfill

\shadowsize 2 pt
\centering
\fboxsep 6 pt{
\shadowbox{
\parbox{\colfaxwidth}{\small
Colfax International (\href{http://www.colfax-intl.com}{http://www.colfax-intl.com/}) is a leading provider of innovative and expertly engineered workstations, servers, clusters, storage, and personal supercomputing solutions. Colfax International is uniquely positioned to offer the broadest spectrum of high performance computing solutions, all of them completely customizable to meet your needs - far beyond anything you can get from any other name brand. Ready-to-go Colfax HPC solutions deliver significant price/performance advantages, and increased IT agility, that accelerates your business and research outcomes. Colfax International's extensive customer base includes Fortune 1000 companies, educational institutions, and government agencies. Founded in 1987, Colfax International is based in Sunnyvale, California and is privately held.
}}}

}

\optcolstart

\printnomenclature

\optcolend
\optcolstart

\section{Introduction}\label{sec:Introduction}

The discrete Fourier transform (DFT) is a universal tool in science and technology.
From image processing \cite{solomon2011fundamentals} to finding distant astronomical objects \cite{2006ApJ...652L..49A}, uses for the Fourier transform are countless and diverse.
Research efforts into efficient numerical Fourier transform algorithms, dubbed discrete fast Fourier transforms (DFFTs), are numerous (e.g., \cite{frigo2005design}) and have led to great improvements in the performance of Fourier transforms in a wide range of configurations on many architectures (e.g., \cite{park2013tera,nukada2012scalable}).

The definition of one-dimensional discrete Fourier transform (1D DFT) is given by the expression
\begin{equation}
\label{eqf}
F_k = \sum\limits_{ n = 0 }^{N-1} x_n \cdot \exp \left[-i\frac{2\pi kn}{N}\right],
\end{equation}
where $F_k$ is the complex $k$-th coefficient in the transform, $x_n$ is the $n$-th data point, and $N$ is the length of the transformed sequence.

If we were to construct a computer algorithm for DFT calculation using Equation~(\ref{eqf}) directly, the number of arithmetic operations (and memory accesses) would scale as $O(N^2)$. 
This quickly becomes ineffective even for small $N$, and for our domain of interest, $N\gtrsim10^9$ (see below), direct application of Equation~(\ref{eqf}) is completely useless. In order to improve the performance, practical implementations of DFTs use optimizations that weaken the complexity scaling and qualify the algorithm as DFFT (discrete {\it fast} Fourier transform).

Most practical applications require either one-dimensional (1D), or multi-dimensional DFFTs with up to a few thousand data values in each dimension. In this paper, we focus on the relatively exotic domain of 1D DFTs of data sets with a few billion data values (i.e., $N\gtrsim 10^9$). This domain of DFTs has application in astronomy (e.g., \cite{2006ApJ...652L..49A}), however, because of its rare occurrence in other fields, existing tools for performing such DFTs are not tuned for multi-core CPUs and perform sub-optimally.

In this paper we describe a new efficient tool for such large 1D DFFTs in shared memory. We call this tool the EFFT library (the acronym stands for ``Enormous Fast Fourier Transforms''). This library is optimized for use on multi-core Intel processors. EFFT uses the parallel framework Intel Cilk Plus in combination with the single-threaded implementation of small DFFTs from the Intel Math Kernel Library (MKL) \cite{mkl}. EFFT performs significantly faster than the two leading DFFT libraries that we tested: FFTW \cite{fftw} and MKL with little increase in memory footprint and with no loss in accuracy. However, unlike those libraries, EFFT supports only 1D transforms of size $N=M\times2^s$, and is optimized only for very large values of $N$.

In addition to the practical importance of the EFFT library, its development process has great educational value. 
That is because the parallel DFFT algorithm is an excellent example of a problem with complex pattern of parallelism.
As such, it is an ideal problem for demonstrating the strength of Intel Cilk Plus framework in effectively parallelizing difficult problems with little need for programmer input.
Furthermore, our implementation of EFFT showcases multiple techniques for optimization for Intel Xeon processors, which are effective for not only DFFT, but a diverse set of algorithms.

In Section~\ref{sec:CTmethod} we discuss the sub-optimal performance of large multi-threaded 1D DFTs in MKL and the possible route to resolution of this inefficiency through the use of the Cooley-Tukey recursive algorithm in combination with parallel framework Intel Cilk Plus. In Section~\ref{sec:EFFTlibrary} we discuss the implementation of EFFT and the optimization consideration that allow us to achieve better performance than MKL. Section~\ref{sec:EFFTdoc} contains brief instructions on using the EFFT library, and Section~\ref{sec:Benchmarks} reports the tuning methodology and performance benchmarks for a range of array sizes on a multi-core processor. Discussion in Section~\ref{sec:Conclusion} touches on the utility of Intel Cilk Plus for complex parallelism and on the applicability of our approach to the Intel Xeon Phi coprocessor architecture.

The result of our work, the EFFT library, is  available for free download at the Colfax Research web site \cite{efft} under the GPLv3 license.

\newpage
\section{Parallelizing the 1D DFFT Calculation}\label{sec:CTmethod}

\subsection{Prior Art: Intel MKL DFT}\label{sec:MKL}

Intel Math Kernel Library (MKL) is a highly optimized, industry-leading mathematics function library. MKL supports DFTs in serial, multi-threaded and cluster implementations. As we are mostly interested in multi-threaded 1D large transforms, in this section we benchmark the MKL performance for this problem domain. 

Listing (\ref{code:dfti}) shows a basic example code that carries out a single-threaded, one-dimensional, real data DFFT in single precision on the data array of the size $N=2^{28}$. 
\begin{listing}[H]
\begin{Verbatim}[commandchars=\\\{\},numbers=left,firstnumber=1,stepnumber=1, ,fontsize=\scriptsize ,frame=single,xleftmargin=10pt,numbersep=2pt]
\PY{c+c1}{// (1) creating the descriptor for single\PYZhy{}}
\PY{c+c1}{//     precision DFFT on real data}
\PY{k}{const} \PY{k+kt}{int} \PY{n}{N} \PY{o}{=} \PY{l+m+mi}{1}\PY{o}{\PYZlt{}\PYZlt{}}\PY{l+m+mi}{28}\PY{p}{;}
\PY{k+kt}{float}\PY{o}{*} \PY{n}{data} \PY{o}{=} \PY{p}{(}\PY{k+kt}{float}\PY{o}{*}\PY{p}{)}
            \PY{n}{\PYZus{}mm\PYZus{}malloc}\PY{p}{(}\PY{n}{N}\PY{o}{*}\PY{k}{sizeof}\PY{p}{(}\PY{k+kt}{float}\PY{p}{),} \PY{l+m+mi}{64}\PY{p}{);}
\PY{n}{MKL\PYZus{}LONG} \PY{n}{fftsize} \PY{o}{=} \PY{n}{N}\PY{p}{;}
\PY{n}{DFTI\PYZus{}DESCRIPTOR\PYZus{}HANDLE}\PY{o}{*} \PY{n}{fftHandle} \PY{o}{=}
                 \PY{k}{new} \PY{n}{DFTI\PYZus{}DESCRIPTOR\PYZus{}HANDLE}\PY{p}{;}
\PY{n}{DftiCreateDescriptor}\PY{p}{(}\PY{n}{fftHandle}\PY{p}{,} \PY{n}{DFTI\PYZus{}SINGLE}\PY{p}{,}
                     \PY{n}{DFTI\PYZus{}REAL}\PY{p}{,} \PY{l+m+mi}{1}\PY{p}{,} \PY{n}{fftsize}\PY{p}{);}

\PY{c+c1}{// (2) configuring the descriptor:}
\PY{c+c1}{//     single\PYZhy{}threaded transform,}
\PY{c+c1}{//     packed permuted data format}
\PY{n}{DftiSetValue}\PY{p}{(}\PY{o}{*}\PY{n}{fftHandle}\PY{p}{,}
            \PY{n}{DFTI\PYZus{}NUMBER\PYZus{}OF\PYZus{}USER\PYZus{}THREADS}\PY{p}{,} \PY{l+m+mi}{1}\PY{p}{);}
\PY{n}{DftiSetValue}\PY{p}{(}\PY{o}{*}\PY{n}{fftHandle}\PY{p}{,} \PY{n}{DFTI\PYZus{}PACKED\PYZus{}FORMAT}\PY{p}{,}
                          \PY{n}{DFTI\PYZus{}PERM\PYZus{}FORMAT}\PY{p}{);}
\PY{n}{DftiCommitDescriptor} \PY{p}{(}\PY{o}{*}\PY{n}{fftHandle}\PY{p}{);}

\PY{c+c1}{// (3) carrying out in\PYZhy{}place DFT}
\PY{n}{DftiComputeForward}\PY{p}{(}\PY{o}{*}\PY{n}{fftHandle}\PY{p}{,} \PY{n}{data}\PY{p}{)}
\end{Verbatim}
\caption{Basic MKL DFT code.\label{code:dfti}}
\end{listing}

The flow of using the DFT functionality in MKL is as follows:
\begin{enumerate}
\tighterenum
\item Create a \texttt{DFTI\_DESCRIPTOR\_HANDLE}. This object stores the parameters of the transform: precision, type, dimensions and size.
\item Set various options for the handle including the data format (layout) and the number of threads. After the change are made, the handle must be finalized with the function \texttt{DftiCommitDescriptor()} before it can be used. Once it has been committed, the handle can be used as many times as the user needs.
\item Carry out the transform using function \texttt{DftiComputeForward(...)}. This runs an in-place forward Fourier transform of the array \texttt{data}.
\end{enumerate}

The format \texttt{DFTI\_PERM\_FORMAT} that we used is an output format that is specific to real-data transforms. With this format, the input data array contains $N$ data points in the following order:
\begin{equation}
\label{eq:dataformat}
\{x_0,\;x_1,\;x_2,\;\dots,\;x_{N-1}\},
\end{equation}
and the output array contains $N/2+1$ complex coefficients $F_k\equiv R_k + i I_k$ in the following order:
\begin{equation}
\label{eq:outputformat}
\{R_{0},R_{\frac{N}{2}},R_{1},I_{1},R_{2},I_{2},\dots,R_{\frac{N}{2}-1},I_{\frac{N}{2}-1}\}.
\end{equation} 
For $k\ge N/2$, the values of $F_k$ can be inferred using the symmetry property expressed below by Equation~(\ref{eqsym}).

Assuming that the user application must process multiple DFTs,
there are two approaches to parallelizing the application:
\begin{enumerate}
\tighterenum
\item Implementing multiple single-threaded Intel MKL DFTs. In this approach, each thread has its own DFTI\_DESCRIPTOR\_HANDLE. The parallelization is done by calling multiple instances of single-threaded \texttt{DftiComputeForward()} from multiple user threads.
\item Taking advantage of the internal threading functionality of the Intel MKL DFT. In this approach, only one DFTI\_DESCRIPTOR\_HANDLE is created, and the value of parameter \texttt{DFTI\_NUMBER\_OF\_USER\_THREADS} is set to the desired number of threads. In this case, the parallelization is done internally by MKL, and multiple cores work in parallel on a single transform.
\end{enumerate}

The first approach is applicable when (a) all DFTs needed by the application are independent from each other, and therefore can be computed concurrently, and (b) the problem size is small enough so that $T$ data arrays can fit in memory, where $T$ is the number of threads used by the application.
The second approach has the advantage of using far less memory, because only one data array must be stored in memory at any given time. This approach also allows to compute each DFT as fast as possible in terms of wall clock time, which is useful when the result of one DFT determines the subsequent operations in the parallel application.

\begin{figure}[H]
\begin{center}
\includegraphics[width=0.5\textwidth, clip=true]{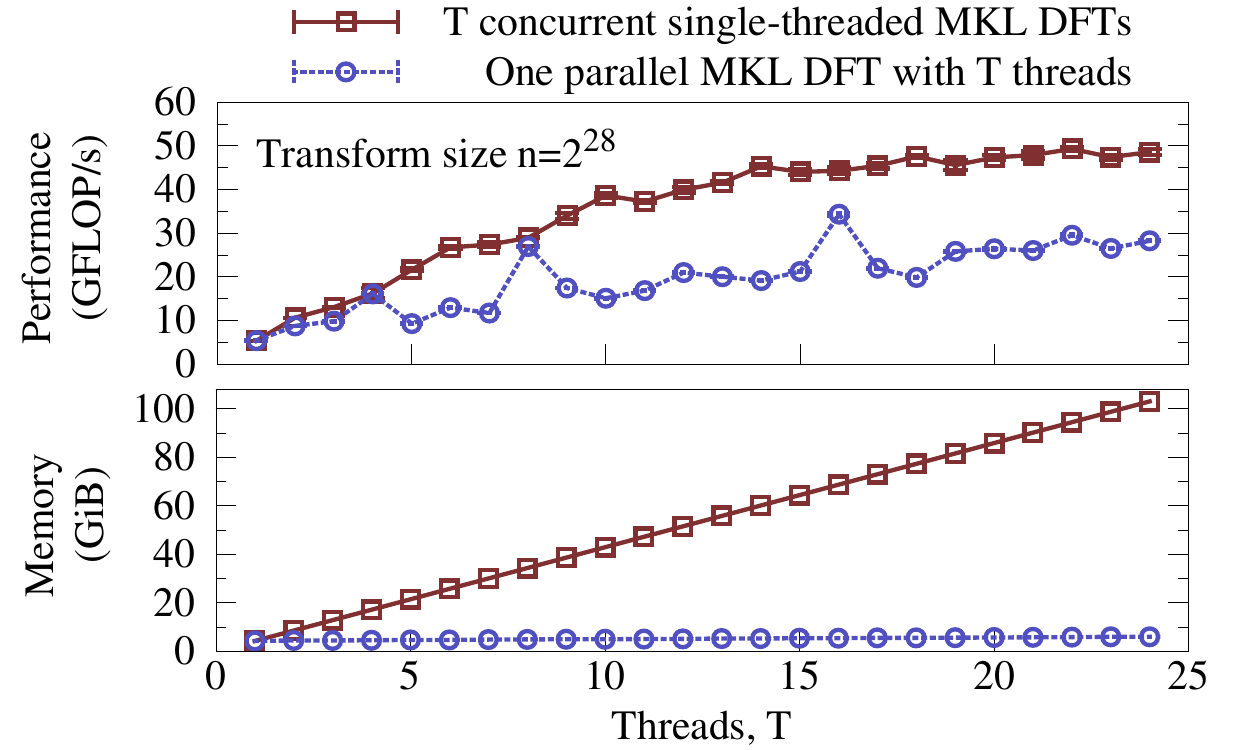}
\caption{{\it Top}: Performance of two parallelization approaches as the function of the number of threads. {\it Bottom}: memory usage in the two approaches. \label{fig:multithreaded} 
}
\end{center}
\end{figure}

Figure (\ref{fig:multithreaded}) shows the performance of the two approaches  as a function of the number of threads for DFTs of size $N=2^{28}$.
The red solid line with square markers reports the performance of multiple single-threaded DFTs, and the dotted blue line with circular markers shows the performance of a single multi-threaded DFT.
In both cases, the total number of threads is set using the \texttt{OMP\_NUM\_THREADS} environment variable. Furthermore in the former case, the number of threads for individual DFT is set to one using \texttt{mkl\_set\_num\_threads()} as well as by setting \texttt{DFTI\_NUMBER\_OF\_USER\_THREADS} in the descriptor handle. 

Based on performance alone, multiple single-threaded DFTs approach has a clear advantage in scalability over the internal threading approach. 
However, running multiple single-threaded DFTs is entirely impractical in the domain of large transforms due to its enormous memory requirement.
The bottom panel of Figure (\ref{fig:multithreaded}) shows the memory usage as a function of number of threads for carrying out Intel MKL DFT.
With multiple concurrent single-threaded DFTs running, the required memory increases linearly with the number of threads $T$.
Thus, in processors with high core count and ($N\gtrsim 10^9$), the memory requirement for the multiple single-threaded DFTs approach may exceed the practical amount of usable RAM.

In this work, we develop the EFFT library for executing multi-threaded DFTs without the performance loss experienced by MKL for large numbers of threads. EFFT builds upon the single-threaded MKL functionality in combination with the Cooley-Tukey algorithm. This algorithm, described in the next section, allows to construct one large DFT from results of many smaller DFTs.

\subsection{Radix-2 Cooley Tukey Algorithm}

The CT algorithm originally presented in \cite{cooley1965algorithm} is the archetype of most DFFT algorithms and is widely used and studied in order to improve the performance of Fourier transforms.
In EFFT, we only use the Radix-2 version, which is a recursive algorithm that breaks down a transform of size $N$ into two smaller transforms of size $N/2$ at each step.
Note that this algorithm requires that the size $N=M\times2^s$, where $s$ is the number of times we intend to apply the recursive splitting.

The core idea of the Radix-2 CT algorithm is to re-write Equation~(\ref{eqf})
as a sum of the elements at even values of $n$ and odd values of $n$:
\begin{eqnarray}
\label{eqb}
F_k &=& \sum\limits_{ n = 0 }^{N/2-1} x_{2n}{\cdot}\exp\left[-i\frac{2\pi k(2n)}{N}\right] + \\
\nonumber
    && + \sum\limits_{ n = 0 }^{N/2-1} x_{2n+1}{\cdot}\exp\left[-i\frac{2\pi k(2n+1)}{N}\right] = \\
\nonumber
 &=& \sum\limits_{ n = 0 }^{N/2-1} x_{2n}{\cdot}\exp\left[-i\frac{2\pi kn}{N/2}\right] + \\
\nonumber
    && + e^{-i2\pi k/N}\sum\limits_{ n = 0 }^{N/2-1} x_{2n+1}{\cdot}\exp\left[-i\frac{2\pi kn}{N/2}\right]\equiv \\
\nonumber
&\equiv& E_k + e^{-i2\pi k/N}O_k.
\end{eqnarray}

By comparing Equation (\ref{eqb}) with (\ref{eqf}), one can see that the first sum is just the Fourier transform of the even-numbered elements and the second sum is the transform of the odd-numbered elements, multiplied by $e^{-i2\pi k/N}$. The multiplication factor $e^{-i2\pi k/N}$ is often referred to as the ``twiddle factor''.
Hereafter, we denote the even part of the transform as $E_k$ and the odd transform as $O_k$.

Values of $E_k$ and $O_k$ need to be found only for $0<k<N/2$, because from the periodicity of the complex exponential functions follows the symmetry

\begin{eqnarray}
\label{eqct}
F_k &=& E_k+e^{-i2\pi k/N}O_k, \\
\label{eqctt}
F_{k+N/2} &=& E_k-e^{-i2\pi k/N}O_k.
\end{eqnarray}
The CT algorithm repeats this decomposition recursively to produce smaller and smaller DFTs.
The number of arithmetic operations (FLOPs) required to compute a transform of size $N$ using the CT recursion is
\begin{equation}
\label{eqflops}
\mathrm{FLOPs} = \frac{5}{2} \times N\log_2N
\end{equation}
This definition of FLOPs conflates the low-latency operations of floating-point addition and multiplication with the long-latency transcendental arithmetic operations, and therefore cannot be easily related to the hardware performance metrics. However, for simplicity, we use this scaling to express the performance of the implementation in terms of GFLOP/s.

For DFTs of real data (i.e., $x_n=x^{*}_n$), the transform coefficients also have the following symmetry;
\begin{equation}
F_k = F_{(N-k-1)}^* \label{eqsym}.
\end{equation}
It is trivial to demonstrate that for all DFTs, $F_0$ and $F_{N/2}$ are both real. 
This means that only coefficients $F_1$ to $F_{N/2-1}$ need to be stored in their full complex form, and the coefficients $F_0$ and $F_{N/2}$ can be stored as real values.
This property allows the result of a real data transform, $F_k$, to be stored in the same sized memory array as the input data $x_n$.

\subsection{Parallel Algorithm of EFFT}

For our purposes, the value of the CT algorithm not only in the reduction of arithmetic complexity from $O(N^2)$ to $O(N\log N)$, but also in its ability to decompose a large Fourier transform into a number of smaller Fourier transforms. 
By decomposing a large transform into many smaller independent transforms, we can execute each of these transforms serially (i.e., with a single processor core) using an existing highly optimized DFT implementation. Parallelism can be achieved by distributing the small serial DFTs across multiple processor cores.

However, for the purposes of optimization, parallelism also has to be achieved in the procedures that precondition the data layout for parallel DFTs. Parallelism is also necessary in the application of Equations~\ref{eqct} and \ref{eqctt}. This makes the parallelization much more challenging.

\begin{figure}[H]
\includegraphics[width=\columnwidth]{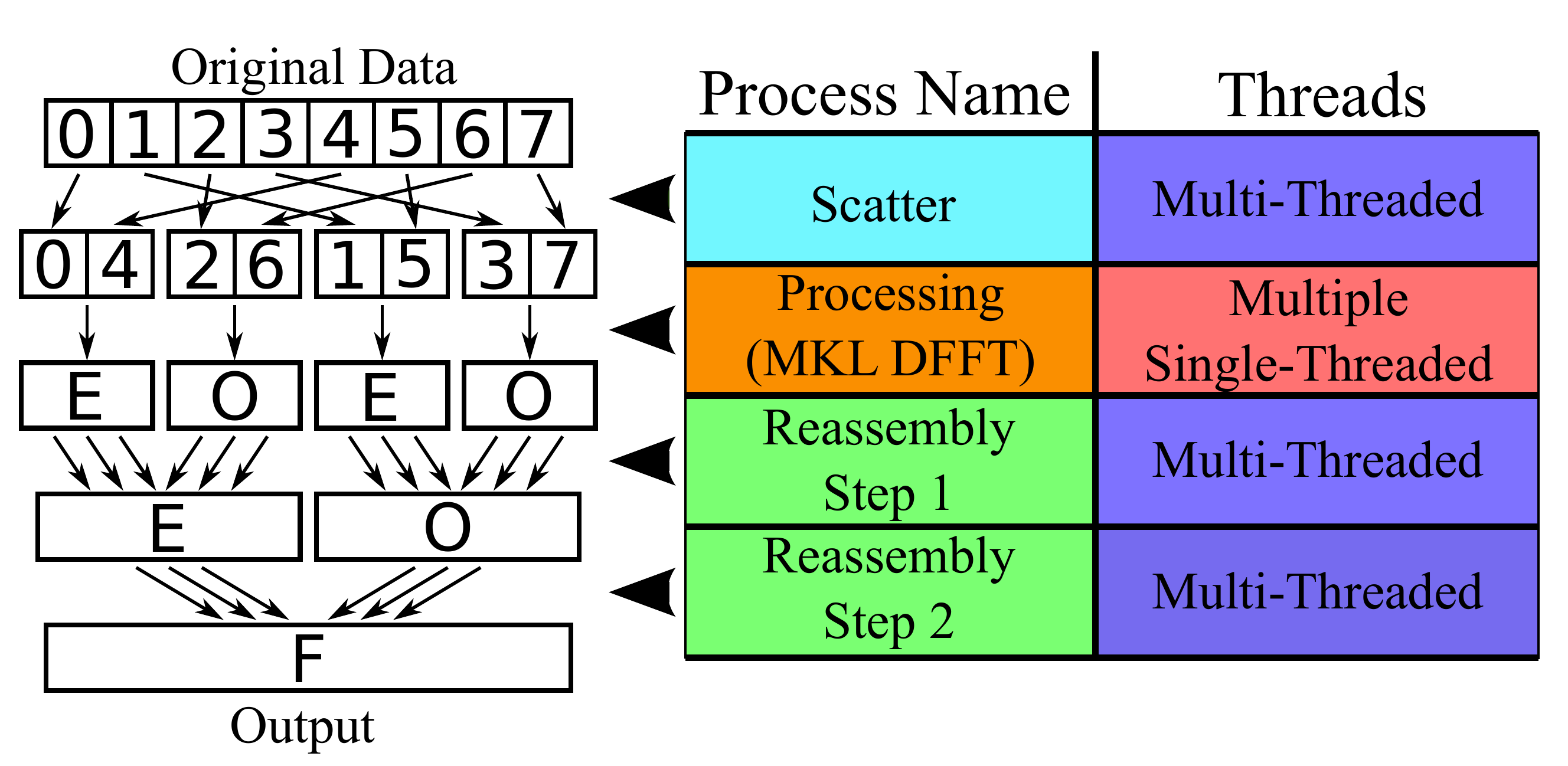}
\caption{Workflow of EFFT with $s=2$ splits and $b=2^s=4$ bins. One  ``scatter'' stage is performed with a multi-threaded loop. It is followed by $b$ single-threaded ``processing'' calls to MKL DFT function. Each pair of consecutive processed bins is ``reassembled'' by a multi-threaded loop. Multiple ``processing'' and ``reassembly'' calls may be running concurrently. \label{fig:stages}}
\end{figure}

In the EFFT library we express the parallel DFFT as three stages:
\begin{enumerate}[I.]
\item ``Scattering'' of $x_n$ into multiple contiguous arrays (``bins''), which are thereafter transformed,
\item ``Processing'' (i.e., performing a serial DFT on) each of the small ``bins'', and
\item ``Reassembly'', i.e., the application of Equations~\ref{eqct} and \ref{eqctt} to the ``bins'' in order to produce the transform coefficients $F_k$.
\end{enumerate}
This workflow is depicted in Figure~\ref{fig:stages}.

Parallelizing the scattering phase across multiple CPU cores is discussed in Section~\ref{sec:Scatter}. 
As for the processing stage, data locality considerations call for a recursive algorithm, as discussed in Section~\ref{sec:processing}. Parallelizing the processing is done by distributing the single-threaded DFTs in each bin across the CPU cores. The difficulty with parallelizing the processing stage is that the number of bins in the simplest algorithm is always a power of 2, however, the number of CPU cores is not necessarily a power of 2. 

Furthermore, data locality considerations suggest that reassembly must be performed as soon as possible after processing. Also, because towards the end of the algorithms, the number of reassembled arrays decreases and eventually gets as low as $2$, scalability considerations require that the reassembly operation itself contain thread parallelism inside.

It can be very difficult in a low-level parallel framework to overlap the multiple independent serial processing tasks with multiple interleaved multi-threaded reassembly tasks, especially if the number of CPU cores is not proportional to the number of tasks. We have tried, but could not come up with a satisfactory implementation using OpenMP, a framework in which the programmer is responsible for assigning specific work items to hardware threads. At the same time, we were able to express the same algorithm in Intel Cilk Plus in a short amount of time, and achieve better results than we were able to obtain with OpenMP. Section~\ref{sec:Cilk} overviews the Intel Cilk Plus framework and explains our implementation.

\subsection{Intel Cilk Plus}\label{sec:Cilk}

Intel Cilk Plus \cite{cilk, cilkweb} is a powerful parallelization and vectorization framework that can effectively parallelize complex problems with very little work required on the part of the programmer.
In this framework, the programmer specifies the components of the application that can be run in parallel, and the runtime library takes care of assigning computing resources (cores) to the parallel tasks.
This is done via an internal scheduling mechanism based on ``work stealing'' to distribute parallel work-items among ``workers''. Workers are a concept in Intel Cilk Plus similar to threads in other frameworks (for example, OpenMP).
Because of the simplicity of the API and high degree of behind-the-scenes automation, Intel Cilk Plus can dramatically reduce the development workload and time while providing great performance.

There are only 3 keywords in Intel Cilk Plus: \texttt{cilk\_for}, \texttt{cilk\_spawn} and \texttt{cilk\_sync}. The keyword \texttt{cilk\_for} is a replacement for the C++ \texttt{for} with a hint that parallel execution is possible in this loop. \texttt{cilk\_spawn} is used to launch a task represented by a function in parallel with the current program. Finally, \texttt{cilk\_sync} is a barrier for all tasks spawned from the current task.
Complex patterns of parallelism, including nested parallel programs and parallel regions requiring different degrees of parallelism, are fully supported.

\begin{listing}[H]
\begin{Verbatim}[commandchars=\\\{\},numbers=left,firstnumber=1,stepnumber=1, ,fontsize=\scriptsize ,frame=single,xleftmargin=10pt,numbersep=2pt]
\PY{n}{cilk\PYZus{}for} \PY{p}{(}\PY{n}{i} \PY{o}{=} \PY{l+m+mi}{0}\PY{p}{;} \PY{n}{i} \PY{o}{\PYZlt{}} \PY{n}{n}\PY{p}{;} \PY{n}{i}\PY{o}{++}\PY{p}{)} \PY{p}{\PYZob{}}
  \PY{c+c1}{// Iterations of this loop}
  \PY{c+c1}{// are distributed between workers}
\PY{p}{\PYZcb{}}

\PY{c+c1}{// Launch MyFunction}
\PY{n}{cilk\PYZus{}spawn} \PY{n}{MyFunction}\PY{p}{(}\PY{n}{a}\PY{p}{,} \PY{n}{b}\PY{p}{,} \PY{n}{c}\PY{p}{);}
\PY{c+c1}{// This will run without waiting}
\PY{c+c1}{// for MyFunction() completion}
\PY{n}{MyOtherFunction}\PY{p}{(}\PY{n}{b}\PY{p}{,} \PY{n}{a}\PY{p}{,} \PY{n}{c}\PY{p}{);}

\PY{c+c1}{// Wait for completion of MyFunction}
\PY{n}{cilk\PYZus{}sync}\PY{p}{;}
\end{Verbatim}
\caption{Summary of Intel Cilk Plus framework keywords.}
\end{listing}

However, note that this ease of use of Cilk functionality does not release the programmer from following the necessary precautions that apply to all parallelized loops.
Avoiding issues such as race conditions is still the programmer's responsibility. 
Intel Cilk Plus provides C++ templates referred to as reducers in order to eliminate race conditions in parallel programs with certain patterns (see, e.g., \cite{mccool2012structured}).

The functionality of Intel Cilk Plus perfectly fits with the parallel pattern of EFFT. Namely, \texttt{cilk\_for} can be used to parallelize the scatter phase and to effect parallelism inside of each reassembly task. \texttt{cilk\_spawn} and \texttt{cilk\_sync} can be used to achieve parallel recursion.

We have limited our discussion of Intel Cilk Plus functionality to those that are relevant to our later discussions.
For complete documentation, visit the Intel C++ Composer Reference Manual \cite{icpc} and the Intel Cilk Plus Web resources \cite{cilk, cilkweb}.

\newpage
\section{Optimizing EFFT}\label{sec:EFFTlibrary}

This section  focuses on the steps we took to parallelize and optimize the EFFT library.

Before we examine the code, we will first describe some of the terminology used in this text and in the code. 
The parameter we call ``number of splits'', represented in codes by variable \texttt{numofsplits}, and in the text by the variable $s$, is the number of \mbox{Radix-2} CT algorithm steps applied to the transform.  
For example, with 3 splits, we apply the Radix-2 CT algorithm 3 times, thus decomposing the transform of size $N$ into $b=2^3=8$ smaller transforms of size $N/8$.
These smaller transforms are stored in segments of a larger ``scratch array''. 
We refer to these segments as ``bins'', which in the code is represented by variable \texttt{numofbins} and in the text by $b$.  
The number of bins is related to the number of splits as $b=2^s$ or $s=\log_2 b$.

\subsection{Initialization}\label{sec:Initialization}

EFFT is a C++ library that implements a class \texttt{EFFT\_Transform}. The constructor of this class takes care of data allocation and initialization for the transform.
Majority of the initialization code is self-explanatory, but we wanted to highlight two important details related to performance optimization.

The first detail is data allocation.
The initialization function in EFFT automatically allocates both the data storage array as well as the necessary temporary storage array. 
The storage array exists as long as the encompassing EFFT class exists.
We retain this array to reduce the overhead of the \texttt{Transform()} method.
Additionally, by allocating the memory internally, we ascertain that (i) the data container is aligned on a 64-byte boundary, and (ii) that the first touch to the data container is performed in a parallel region by the MKL DFT initialization function. Both of these details are important for subsequent performance of the FFT.

The second important aspect of EFFT initialization is MKL DFT handle creation. 
Since the time required to create a handle is often longer than the actual transform itself, it is important to create the handle once and retain it for all subsequent transforms. 
Furthermore, in order to implement the most efficient MKL usage mode, each Intel Cilk Plus worker creates its own handle to be used in all subsequent transforms.

Listing (\ref{code:init}) shows the portions of the initialization that we have discussed.
\begin{listing}[H]
\begin{Verbatim}[commandchars=\\\{\},numbers=left,firstnumber=1,stepnumber=1, ,fontsize=\scriptsize ,frame=single,xleftmargin=10pt,numbersep=2pt]
\PY{c+c1}{// Data allocation on 64 byte boundary}
\PY{n}{dataPtr} \PY{o}{=}
   \PY{p}{(}\PY{k+kt}{float}\PY{o}{*}\PY{p}{)}\PY{n}{\PYZus{}mm\PYZus{}malloc}\PY{p}{(}\PY{n}{size}\PY{o}{*}\PY{k}{sizeof}\PY{p}{(}\PY{k+kt}{float}\PY{p}{),} \PY{l+m+mi}{64}\PY{p}{);}
\PY{n}{scratchPtr} \PY{o}{=}
  \PY{p}{(}\PY{k+kt}{float}\PY{o}{*}\PY{p}{)} \PY{n}{\PYZus{}mm\PYZus{}malloc}\PY{p}{(}\PY{n}{size}\PY{o}{*}\PY{k}{sizeof}\PY{p}{(}\PY{k+kt}{float}\PY{p}{),} \PY{l+m+mi}{64}\PY{p}{);}

\PY{c+c1}{// Handle creation}
\PY{n}{DFTI\PYZus{}DESCRIPTOR\PYZus{}HANDLE}\PY{o}{*} \PY{n}{fftHandle} \PY{o}{=}
 \PY{p}{(}\PY{n}{DFTI\PYZus{}DESCRIPTOR\PYZus{}HANDLE}\PY{o}{*}\PY{p}{)}\PY{n}{malloc}\PY{p}{(}\PY{n}{numthreads}\PY{o}{*}
              \PY{k}{sizeof}\PY{p}{(}\PY{n}{DFTI\PYZus{}DESCRIPTOR\PYZus{}HANDLE}\PY{p}{));}

\PY{c+c1}{// Spawning a parallel region}
\PY{n}{cilk\PYZus{}for} \PY{p}{(}\PY{k+kt}{int} \PY{n}{i} \PY{o}{=} \PY{l+m+mi}{0}\PY{p}{;} \PY{n}{i} \PY{o}{\PYZlt{}} \PY{n}{numworkers}\PY{p}{;} \PY{n}{i}\PY{o}{++}\PY{p}{)\PYZob{}}
  \PY{k+kt}{int} \PY{n}{wid} \PY{o}{=} \PY{n}{\PYZus{}\PYZus{}cilkrts\PYZus{}get\PYZus{}worker\PYZus{}number}\PY{p}{();}
  \PY{n}{mkl\PYZus{}set\PYZus{}num\PYZus{}threads}\PY{p}{(}\PY{l+m+mi}{1}\PY{p}{);}
  \PY{c+c1}{// Preparing small MKL FFT}
  \PY{n}{MKL\PYZus{}LONG} \PY{n}{fftsize} \PY{o}{=} \PY{n}{binsize}\PY{p}{;}
  \PY{c+c1}{// One handle per worker}
  \PY{n}{DftiCreateDescriptor}\PY{p}{(}\PY{o}{\PYZam{}}\PY{n}{fftHandle}\PY{p}{[}\PY{n}{wid}\PY{p}{],}
          \PY{n}{DFTI\PYZus{}SINGLE}\PY{p}{,} \PY{n}{DFTI\PYZus{}REAL}\PY{p}{,} \PY{l+m+mi}{1}\PY{p}{,} \PY{n}{fftsize}\PY{p}{);}
  \PY{n}{DftiSetValue}\PY{p}{(}\PY{n}{fftHandle}\PY{p}{[}\PY{n}{wid}\PY{p}{],}
              \PY{n}{DFTI\PYZus{}NUMBER\PYZus{}OF\PYZus{}USER\PYZus{}THREADS}\PY{p}{,} \PY{l+m+mi}{1}\PY{p}{);}
  \PY{n}{DftiSetValue}\PY{p}{(}\PY{n}{fftHandle}\PY{p}{[}\PY{n}{wid}\PY{p}{],}
        \PY{n}{DFTI\PYZus{}PACKED\PYZus{}FORMAT}\PY{p}{,} \PY{n}{DFTI\PYZus{}PERM\PYZus{}FORMAT}\PY{p}{);}
  \PY{n}{DftiCommitDescriptor} \PY{p}{(}\PY{n}{fftHandle}\PY{p}{[}\PY{n}{wid}\PY{p}{]);}
\PY{p}{\PYZcb{}}
\end{Verbatim}
\vskip -0.5em
\caption{Data allocation and handle creation a initialization.\label{code:init}}
\end{listing}

\subsection{Scattering Stage}\label{sec:Scatter}

The purpose of the scatter phase of the transform is to reorganize the data set into a temporary data array, where the elements for small DFTs are listed contiguously.
In terms of the CT algorithm, this is the phase where we separate the even-numbered elements and odd-numbered elements. 

Listing (\ref{code:basescatter}) shows a basic, unoptimized scatter that is equivalent to one step in the Radix-2 CT algorithm.
\begin{listing}[H]
\begin{Verbatim}[commandchars=\\\{\},numbers=left,firstnumber=1,stepnumber=1, ,fontsize=\scriptsize ,frame=single,xleftmargin=10pt,numbersep=2pt]
\PY{k}{for} \PY{p}{(}\PY{k+kt}{int} \PY{n}{i} \PY{o}{=} \PY{l+m+mi}{0}\PY{p}{;} \PY{n}{i} \PY{o}{\PYZlt{}} \PY{n}{N}\PY{p}{;} \PY{n}{i}\PY{o}{+=}\PY{l+m+mi}{2}\PY{p}{)} \PY{p}{\PYZob{}}
  \PY{n}{temparray}\PY{p}{[}\PY{n}{i}\PY{p}{]} \PY{o}{=} \PY{n}{data}\PY{p}{[}\PY{l+m+mi}{2}\PY{o}{*}\PY{n}{i}\PY{p}{];}
  \PY{n}{temparray}\PY{p}{[}\PY{n}{i}\PY{o}{+}\PY{n}{N}\PY{o}{/}\PY{l+m+mi}{2}\PY{p}{]} \PY{o}{=} \PY{n}{data}\PY{p}{[}\PY{l+m+mi}{2}\PY{o}{*}\PY{n}{i}\PY{o}{+}\PY{l+m+mi}{1}\PY{p}{];}
\PY{p}{\PYZcb{}}
\end{Verbatim}
\vskip -0.5em
\caption{Basic recursive scatter.\label{code:basescatter}}
\end{listing}
This basic version of scatter can be applied recursively to arrive at the reorganized array we require, but has extremely poor performance due to high number of memory accesses.
In order to carry out the Radix-2 CT algorithm $s$ times, or to do $s$ splits, this recursive scatter will have to read $2 \cdot s \cdot N$ memory addresses. 
Thus, in order to gain performance it would be ideal to do the full scatter in a single pass.

In our EFFT code, we use the fact that $s$ is generally small and precompute the pattern of mapping from the original data array to the scratch array. This mapping is based on the bit reversal principle\cite{cooley1965algorithm}.
Listing (\ref{code:unoptimizedscatter}) shows a (still unoptimized) variant of our that version.
In that code, we also use $cilk\_for$ to parallelize the scattering process.

\begin{listing}[H]
\begin{Verbatim}[commandchars=\\\{\},numbers=left,firstnumber=1,stepnumber=1, ,fontsize=\scriptsize ,frame=single,xleftmargin=10pt,numbersep=2pt]
\PY{k+kt}{int} \PY{n}{scatter\PYZus{}index}\PY{p}{[}\PY{n}{numofbins}\PY{p}{];}
\PY{n}{find\PYZus{}scatter\PYZus{}index}\PY{p}{(}\PY{n}{scatter\PYZus{}index}\PY{p}{,}\PY{n}{numofsplits}\PY{p}{);}
\PY{n}{cilk\PYZus{}for}\PY{p}{(}\PY{k+kt}{int} \PY{n}{j} \PY{o}{=} \PY{l+m+mi}{0}\PY{p}{;} \PY{n}{j} \PY{o}{\PYZlt{}} \PY{n}{numofbins}\PY{p}{;} \PY{n}{j}\PY{o}{++}\PY{p}{)}
  \PY{k}{for} \PY{p}{(}\PY{k+kt}{long} \PY{n}{i} \PY{o}{=} \PY{l+m+mi}{0}\PY{p}{;} \PY{n}{i} \PY{o}{\PYZlt{}} \PY{n}{binsize}\PY{p}{;} \PY{n}{i}\PY{o}{++}\PY{p}{)}
    \PY{n}{tempArray}\PY{p}{[}\PY{n}{scatter\PYZus{}index}\PY{p}{[}\PY{n}{j}\PY{p}{]}\PY{o}{*}\PY{n}{binsize}\PY{o}{+}\PY{n}{i}\PY{p}{]} \PY{o}{=}
                     \PY{n}{dataArray}\PY{p}{[}\PY{n}{i}\PY{o}{*}\PY{n}{numofbins}\PY{o}{+}\PY{n}{j}\PY{p}{];}
\end{Verbatim}
\vskip -0.5em
\caption{Unoptimized scatter.\label{code:unoptimizedscatter}}
\end{listing}

There are several techniques we can employ to optimize this scatter code for parallelism and improve its performance.
First technique is to permute the $j$ and the $i$-loops. This is helpful because in a target parameter domain of EFFT ($N\gtrsim 10^9$), practical values of $b\equiv$~\texttt{numofbins} are much smaller than $N/s\equiv$~\texttt{binsize}.
Therefore, in modern multi-core processors, parallelizing the $j$-loop with $b$ iterations can potentially lead to the problem of not having enough parallelism.
For instance, if we had $T=24$ workers and $b=32$ bins, the runtime system would first distribute 24 bins across the 24 workers. Once this work is done, only $32-24=8$ of the 24 workers will be utilized for the remainder of the calculation. 
Therefore, permuting the loops to bring the $i$ loop outside and parallelze it ensures that there is enough parallelism in the code. 

However, the loop permutation has the negative effect of losing unit-stride access in $i$ in the inner loop.
It is true that we gain unit-stride in $j$ in exchange for this loss, but for this particular implementation of the scatter, unit-stride in $i$ is of greater importance. 
This is due to the fact that the variable $i$ is multiplied by $b\equiv$~\texttt{numofbins}, while $j$ (or, rather, \texttt{scatter\_index[j]}) is multiplied by \texttt{binsize}~$\equiv N/b$.
Again, the $b$ is normally much smaller than $N/s$, so memory accesses with consecutive values of $i$ are located closer to each other than accesses consecutive in $j$. 
Thus having $i$ in the inner loop is more optimal.

In order to get parallelism in $i$ while keeping a unit-stride access in $i$, we implement the technique called loop tiling.
This technique involves strip-mining the inner loop and permuting the outer two loops.
The strip-mining operation expresses a single loop as two nested loops, of which the outer loop iterates with a stride referred to as ``Tile'', and the inner loop iterates incrementally within the ``Tile''.
Listing (\ref{code:scatter}) shows our final optimized code.
The value \texttt{iTile=16} is obtained empirically and provides the best performance in most cases.
 
\begin{listing}[H]
\begin{Verbatim}[commandchars=\\\{\},numbers=left,firstnumber=1,stepnumber=1, ,fontsize=\scriptsize ,frame=single,xleftmargin=10pt,numbersep=2pt]
\PY{k+kt}{int} \PY{n}{scatter\PYZus{}index}\PY{p}{[}\PY{n}{numofbins}\PY{p}{];}
\PY{n}{find\PYZus{}scatter\PYZus{}index}\PY{p}{(}\PY{n}{scatter\PYZus{}index}\PY{p}{,} \PY{n}{numofsplits}\PY{p}{);}
\PY{k}{const} \PY{k+kt}{long} \PY{n}{iTILE}\PY{o}{=}\PY{l+m+mi}{16}\PY{p}{;}
\PY{k+kt}{long} \PY{n}{ii}\PY{p}{;}
\PY{n}{cilk\PYZus{}for} \PY{p}{(}\PY{n}{ii} \PY{o}{=} \PY{l+m+mi}{0}\PY{p}{;} \PY{n}{ii} \PY{o}{\PYZlt{}} \PY{n}{binsize}\PY{p}{;} \PY{n}{ii} \PY{o}{+=} \PY{n}{iTILE}\PY{p}{)}
  \PY{k}{for}\PY{p}{(}\PY{k+kt}{long} \PY{n}{j} \PY{o}{=} \PY{l+m+mi}{0}\PY{p}{;} \PY{n}{j} \PY{o}{\PYZlt{}} \PY{n}{numofbins}\PY{p}{;} \PY{n}{j}\PY{o}{++}\PY{p}{)}
    \PY{k}{for} \PY{p}{(}\PY{k+kt}{long} \PY{n}{i} \PY{o}{=} \PY{n}{ii}\PY{p}{;} \PY{n}{i} \PY{o}{\PYZlt{}} \PY{n}{ii}\PY{o}{+}\PY{n}{iTILE}\PY{p}{;} \PY{n}{i}\PY{o}{++}\PY{p}{)}
      \PY{n}{tempArray}\PY{p}{[}\PY{n}{scatter\PYZus{}index}\PY{p}{[}\PY{n}{j}\PY{p}{]}\PY{o}{*}\PY{n}{binsize}\PY{o}{+}\PY{n}{i}\PY{p}{]} \PY{o}{=}
                     \PY{n}{dataArray}\PY{p}{[}\PY{n}{i}\PY{o}{*}\PY{n}{numofbins}\PY{o}{+}\PY{n}{j}\PY{p}{];}
\end{Verbatim}
\vskip -0.5em
\caption{Optimized scatter code with tiling.\label{code:scatter}}
\end{listing}

In order to understand why Listing~\ref{code:scatter} is optimal, consider the locality of data accesses.
Each worker operates on a different value of \texttt{ii} (in practice, each worker will get a contiguous chunk of values of \texttt{ii} to process).
For each value of \texttt{ii} and each value of \texttt{j}, the worker reads \texttt{iTile==16} elements from \texttt{dataArray} with a stride of \texttt{numofbins}. Then these elements are written contiguously into \texttt{tempArray}. Contiguous access to \texttt{tempArray} is good, however, in the process of reading \texttt{dataArray}, the worker had touch 16 cache lines to read only 16 words. Luckily, the loop in \texttt{i} is short enough so that for the next value of $j$, these cache lines are re-used. Therefore, the application can achieve a significant fraction of memory bandwidth. At the same time, the outer parallel loop has enough iterations to scale across tens of threads available in modern CPUs.

\subsection{Processing/Parallel Recursion}\label{sec:processing}

In order to parallelize the processing and reassembly phases, we have combined them into a single parallel recursion tree. This framework is shown in Listing~\ref{code:parallelrecursion}.

\begin{listing}[H]
\begin{Verbatim}[commandchars=\\\{\},numbers=left,firstnumber=1,stepnumber=1, ,fontsize=\scriptsize ,frame=single,xleftmargin=10pt,numbersep=2pt]
\PY{k+kt}{void} \PY{n+nf}{ProcessAndReassemble}\PY{p}{(}
         \PY{n}{DFTI\PYZus{}DESCRIPTOR\PYZus{}HANDLE}\PY{o}{*} \PY{n}{fftHandle}\PY{p}{,}
         \PY{k+kt}{float}\PY{o}{*} \PY{n}{array}\PY{p}{,} \PY{k+kt}{float}\PY{o}{*} \PY{n}{temp}\PY{p}{,}
         \PY{k}{const} \PY{k+kt}{long} \PY{n}{n}\PY{p}{,} \PY{k}{const} \PY{k+kt}{long} \PY{n}{binsize}\PY{p}{)} \PY{p}{\PYZob{}}
  \PY{k}{const} \PY{k+kt}{long} \PY{n}{size} \PY{o}{=} \PY{n}{n}\PY{o}{/}\PY{l+m+mi}{2}\PY{p}{;}

  \PY{k}{if} \PY{p}{(}\PY{n}{n} \PY{o}{\PYZgt{}} \PY{n}{binsize}\PY{p}{)} \PY{p}{\PYZob{}}
    \PY{n}{cilk\PYZus{}spawn} \PY{c+c1}{// Create parallel task}
      \PY{n}{ProcessAndReassemble}\PY{p}{(}\PY{n}{fftHandle}\PY{p}{,}
                \PY{n}{array}\PY{p}{,} \PY{n}{temp}\PY{p}{,} \PY{n}{size}\PY{p}{,} \PY{n}{binsize}\PY{p}{);}
    \PY{n}{ProcessAndReassemble}\PY{p}{(}\PY{n}{fftHandle}\PY{p}{,}
                \PY{o}{\PYZam{}}\PY{n}{array}\PY{p}{[}\PY{n}{size}\PY{p}{],} \PY{o}{\PYZam{}}\PY{n}{temp}\PY{p}{[}\PY{n}{size}\PY{p}{],}
                             \PY{n}{size}\PY{p}{,} \PY{n}{binsize}\PY{p}{);}
    \PY{n}{cilk\PYZus{}sync}\PY{p}{;} \PY{c+c1}{// Wait for spawned task}

    \PY{c+c1}{// CT algorithm to combine evens and odds}
    \PY{c+c1}{// ReassemblePair() is multithreaded}
    \PY{n}{ReassemblePair}\PY{p}{(}\PY{n}{temp}\PY{p}{,} \PY{n}{size}\PY{p}{);}
  \PY{p}{\PYZcb{}} \PY{k}{else} \PY{p}{\PYZob{}}
    \PY{c+c1}{// Process the small enough array}
    \PY{c+c1}{// with serial MKL implementation of DFT}
    \PY{k+kt}{int} \PY{n}{wid} \PY{o}{=} \PY{n}{\PYZus{}\PYZus{}cilkrts\PYZus{}get\PYZus{}worker\PYZus{}number}\PY{p}{();}
    \PY{n}{DftiComputeForward}\PY{p}{(}\PY{n}{fftHandle}\PY{p}{[}\PY{n}{wid}\PY{p}{],} \PY{n}{temp}\PY{p}{);}
  \PY{p}{\PYZcb{}}
\PY{p}{\PYZcb{}}
\end{Verbatim}
\caption{Parallel recursive processing and reassembly in EFFT.\label{code:parallelrecursion}}
\end{listing}

The processing call is initially applied to the array of size \texttt{n=N}. This call subsequently recurses into two instances of itself with \texttt{n} reduced by a factor of 2. One of the recursive calls is placed in an Intel Cilk Plus task using the keyword \texttt{cilk\_spawn}, thus effecting parallel recursion. Recursion stops then \texttt{n} is reduced to the target \texttt{binsize}, and processing with serial MKL DFT is applied in parallel to the multiple small ``bins''. After the processing of two adjacent bins is complete (as indicated by the return of the \texttt{cilk\_sync} call), the reassembly function is called on these two bins. Reassembly has thread parallelism inside, as explained in Section~\ref{sec:reassembly}

To see why we chose to incorporate reassembly in the same parallel recursion tree as processing, consider a case where the architecture allows for $T=24$ threads, and we do $s=5$ splits to get $b=32$ bins.
The code starts the first 24 serial DFTs in parallel, as desired.
However, after the first 24 bins are processed (transformed), the rest of the DFTs can occupy only 8 threads.
In order to utilize all available CPU resources, the remaining 16 threads should work on reassembly while the other 8 finish the rest of the DFTs.
Thus, in order to utilize all available hardware in a parallel setting, it is beneficial to combine reassembly and processing in the same tree.

Implementing a parallel framework to execute such a parallel algorithm is not trivial.
One of the complications is that the number of workers available for the reassembly varies as DFTs get finished.
Keeping track of how many threads are available in a parallel region is complicated, as it requires monitoring which bins can be assembled and by which thread. 
Another issue is that multiple calls to multi-threaded reassembly function may run simultaneously. 
Controlling the order of their execution and preventing the system from over-subscription of threads can be challenging.
In cases like this, the Intel Cilk Plus framework can be a convenient solution.
Work items are assigned to workers automatically through the runtime library's high performing scheduling algorithm.

\subsection{Reassembly}\label{sec:reassembly}

The purpose of the reassembly phase is to apply Equations~(\ref{eqct}) and (\ref{eqctt}) to assemble the results of smaller DFTs to produce the result of the larger decomposed FFT. This stage is performed by the function \texttt{ReassemblePair()} called from \texttt{ProcessAndReassemble()} in Listing~\ref{code:parallelrecursion}.

The reassembly component uses the Radix-2 CT algorithm to recombine the results of the smaller DFTs. 
Listing~\ref{code:baseassembly} shows a basic (i.e., unoptimized) implementation of reassembly that applies a singe step of Radix-2 CT algorithm to combine two bins into one.

In Listing~\ref{code:baseassembly}, \texttt{evens} and \texttt{odds} are the FFT result of even- and odd-numbered elements, respectively ($E_k$ and $O_k$), \texttt{target} is the target location to stored the combined result ($F_k$), and $size$ is the size of the \texttt{evens} and \texttt{odds} arrays.
Note that the first two elements are dealt with as special cases, because those are the real parts of $F_0$ and $F_{N/2}$ instead of the real-imaginary components of $F_{k}\equiv R_k + iI_k$ like the rest (see Equation~\ref{eq:outputformat}).

\begin{listing}[H]
\begin{Verbatim}[commandchars=\\\{\},numbers=left,firstnumber=1,stepnumber=1, ,fontsize=\scriptsize ,frame=single,xleftmargin=10pt,numbersep=2pt]
\PY{k+kt}{void} \PY{n+nf}{BasicReassemble}\PY{p}{(}\PY{k+kt}{float}\PY{o}{*} \PY{n}{evens}\PY{p}{,} \PY{k+kt}{float}\PY{o}{*} \PY{n}{odds}\PY{p}{,}
         \PY{k+kt}{float}\PY{o}{*} \PY{n}{target}\PY{p}{,} \PY{k}{const} \PY{k+kt}{long} \PY{k+kt}{int} \PY{n}{size}\PY{p}{)} \PY{p}{\PYZob{}}
  \PY{k}{const} \PY{k+kt}{float} \PY{n}{trigconst} \PY{o}{=} \PY{o}{\PYZhy{}}\PY{l+m+mf}{3.14159265359f}\PY{o}{/}\PY{n}{size}\PY{p}{;}
  \PY{n}{target}\PY{p}{[}\PY{l+m+mi}{0}      \PY{p}{]}\PY{o}{=}\PY{n}{evens}\PY{p}{[}\PY{l+m+mi}{0}\PY{p}{]} \PY{o}{+}\PY{n}{odds}\PY{p}{[}\PY{l+m+mi}{0}\PY{p}{];}
  \PY{n}{target}\PY{p}{[}\PY{l+m+mi}{1}      \PY{p}{]}\PY{o}{=}\PY{n}{evens}\PY{p}{[}\PY{l+m+mi}{0}\PY{p}{]} \PY{o}{\PYZhy{}}\PY{n}{odds}\PY{p}{[}\PY{l+m+mi}{0}\PY{p}{];}
  \PY{n}{target}\PY{p}{[}\PY{n}{size}   \PY{p}{]}\PY{o}{=}\PY{n}{evens}\PY{p}{[}\PY{l+m+mi}{1}\PY{p}{];}
  \PY{n}{target}\PY{p}{[}\PY{n}{size}\PY{o}{+}\PY{l+m+mi}{1L}\PY{p}{]}\PY{o}{=\PYZhy{}}\PY{n}{odds}\PY{p}{[}\PY{l+m+mi}{1}\PY{p}{];}

  \PY{k}{for}\PY{p}{(}\PY{k+kt}{long} \PY{k+kt}{int} \PY{n}{k} \PY{o}{=} \PY{l+m+mi}{1L}\PY{p}{;} \PY{n}{k} \PY{o}{\PYZlt{}} \PY{n}{size}\PY{o}{/}\PY{l+m+mi}{2L}\PY{p}{;} \PY{n}{k}\PY{o}{++}\PY{p}{)} \PY{p}{\PYZob{}}
    \PY{k+kt}{float} \PY{n}{cosk} \PY{o}{=} \PY{n}{cosf}\PY{p}{(}\PY{n}{k}\PY{o}{*}\PY{n}{trigconst}\PY{p}{);}
    \PY{k+kt}{float} \PY{n}{sink} \PY{o}{=} \PY{n}{sinf}\PY{p}{(}\PY{n}{k}\PY{o}{*}\PY{n}{trigconst}\PY{p}{);}
    \PY{n}{target}\PY{p}{[}\PY{l+m+mi}{2}\PY{o}{*}\PY{n}{k}\PY{p}{]}\PY{o}{=}\PY{n}{evens}\PY{p}{[}\PY{l+m+mi}{2}\PY{o}{*}\PY{n}{k}\PY{p}{]}\PY{o}{+}\PY{n}{odds}\PY{p}{[}\PY{l+m+mi}{2}\PY{o}{*}\PY{n}{k}\PY{p}{]}\PY{o}{*}\PY{n}{cosk}
                          \PY{o}{\PYZhy{}}\PY{n}{odds}\PY{p}{[}\PY{l+m+mi}{2}\PY{o}{*}\PY{n}{k}\PY{o}{+}\PY{l+m+mi}{1}\PY{p}{]}\PY{o}{*}\PY{n}{sink}\PY{p}{;}
    \PY{n}{target}\PY{p}{[}\PY{l+m+mi}{2}\PY{o}{*}\PY{n}{k}\PY{o}{+}\PY{l+m+mi}{1}\PY{p}{]}\PY{o}{=}\PY{n}{evens}\PY{p}{[}\PY{l+m+mi}{2}\PY{o}{*}\PY{n}{k}\PY{o}{+}\PY{l+m+mi}{1}\PY{p}{]}\PY{o}{+}\PY{n}{odds}\PY{p}{[}\PY{l+m+mi}{2}\PY{o}{*}\PY{n}{k}\PY{p}{]}\PY{o}{*}\PY{n}{sink}
                            \PY{o}{+}\PY{n}{odds}\PY{p}{[}\PY{l+m+mi}{2}\PY{o}{*}\PY{n}{k}\PY{o}{+}\PY{l+m+mi}{1}\PY{p}{]}\PY{o}{*}\PY{n}{cosk}\PY{p}{;}
    \PY{n}{target}\PY{p}{[}\PY{l+m+mi}{2}\PY{o}{*}\PY{n}{size}\PY{o}{\PYZhy{}}\PY{l+m+mi}{2}\PY{o}{*}\PY{n}{k}\PY{p}{]}\PY{o}{=}\PY{n}{evens}\PY{p}{[}\PY{l+m+mi}{2}\PY{o}{*}\PY{n}{k}\PY{p}{]}
             \PY{o}{\PYZhy{}}\PY{n}{odds}\PY{p}{[}\PY{l+m+mi}{2}\PY{o}{*}\PY{n}{k}\PY{p}{]}\PY{o}{*}\PY{n}{cosk}\PY{o}{+}\PY{n}{odds}\PY{p}{[}\PY{l+m+mi}{2}\PY{o}{*}\PY{n}{k}\PY{o}{+}\PY{l+m+mi}{1}\PY{p}{]}\PY{o}{*}\PY{n}{sink}\PY{p}{;}
    \PY{n}{target}\PY{p}{[}\PY{l+m+mi}{2}\PY{o}{*}\PY{n}{size}\PY{o}{\PYZhy{}}\PY{l+m+mi}{2}\PY{o}{*}\PY{n}{k}\PY{o}{+}\PY{l+m+mi}{1}\PY{p}{]}\PY{o}{=\PYZhy{}}\PY{n}{evens}\PY{p}{[}\PY{l+m+mi}{2}\PY{o}{*}\PY{n}{k}\PY{o}{+}\PY{l+m+mi}{1}\PY{p}{]}
             \PY{o}{+}\PY{n}{odds}\PY{p}{[}\PY{l+m+mi}{2}\PY{o}{*}\PY{n}{k}\PY{p}{]}\PY{o}{*}\PY{n}{sink}\PY{o}{+}\PY{n}{odds}\PY{p}{[}\PY{l+m+mi}{2}\PY{o}{*}\PY{n}{k}\PY{o}{+}\PY{l+m+mi}{1}\PY{p}{]}\PY{o}{*}\PY{n}{cosk}\PY{p}{;}
  \PY{p}{\PYZcb{}}
\PY{p}{\PYZcb{}}
\end{Verbatim}
\vskip -0.5em
\caption{Basic (not optimized) reassembly of two arrays into one using Equation~(\ref{eqct}). \label{code:baseassembly}}
\end{listing}

Following equation (\ref{eqct}), the code adds the $k$-th element of the \texttt{evens} to the k-th element of the \texttt{odds} multiplied by the twiddle factor.
The second half of the target array, which corresponds to  $k\geq N/2$ (which are not in the \texttt{evens} and \texttt{odds} arrays), is calculated by taking advantage the symmetry expressed by Equation~(\ref{eqsym}).

At first glance this appears to be a problem that is easily vectorized, since the result is simply a superposition of elements of two arrays multiplied by trigonometric functions, which can be vectorized.
However, because the arrays contain complex numbers with real and imaginary parts interleaved, access to the potentially vectorizable data has a stride of 2.
This situation is difficult for the compiler to handle, and we have found that the following code modification improves automatic vectorization and, along with it, application performance.

First, in order to vectorize the calculation of trigonometric functions, we strip-mine the array in $k$. This operation expresses the loop in $k$ as two nested loops in $kk$ (iterating with a stride of 64) and a loop in $k$ (iterating with a stride of 1 through 64 iterations). The second step is un-fusing the loop over the trigonometric functions from the loop over array elements (see Listing~\ref{code:stripminedreassembly}). The reason for strip-mining the loop is to restrict the size of the container for precomputed trigonometric functions to a small enough value that does not cause the eviction of \texttt{evens} and \texttt{odds} from caches.
Note that with the strip-mined loop, 
we have to handle $k=0$ as an exception (see output format given by  Equation~\ref{eq:outputformat}).
This also necessitates a separate loop for processing elements from $k=1$ to $k=kTILE-1$, because the automatically vectorized loops in lines 18 and 24 of Listing~\ref{code:stripminedreassembly} is built for complete tiles of size $kTILE$. Separate loop for the first tile introduces redundant code, however, it allows us to avoid branches in the main loops in lines 18 and 24, which would ruin the overall performance of the application.

\begin{listing}[H]
\begin{Verbatim}[commandchars=\\\{\},numbers=left,firstnumber=1,stepnumber=1, ,fontsize=\scriptsize ,frame=single,xleftmargin=10pt,numbersep=2pt]
\PY{k}{const} \PY{k+kt}{long} \PY{n}{kTILE} \PY{o}{=} \PY{l+m+mi}{64L}\PY{p}{;}

\PY{c+c1}{// ...omitted the code handling 0 and N/2}

\PY{k}{for}\PY{p}{(}\PY{k+kt}{long} \PY{n}{k} \PY{o}{=} \PY{l+m+mi}{1}\PY{p}{;} \PY{n}{k} \PY{o}{\PYZlt{}} \PY{n}{kTILE}\PY{p}{;} \PY{n}{kk}\PY{o}{+=}\PY{n}{kTILE}\PY{p}{)} \PY{p}{\PYZob{}}
  \PY{c+c1}{// Elements from 1 to kTILE\PYZhy{}1}
  \PY{c+c1}{// are handled separately}
\PY{p}{\PYZcb{}}

\PY{c+cp}{\PYZsh{}define ALIGNED \PYZus{}\PYZus{}attribute\PYZus{}\PYZus{}((aligned(64)))}
\PY{n}{cilk\PYZus{}for}\PY{p}{(}\PY{k+kt}{long} \PY{n}{kk} \PY{o}{=} \PY{n}{kTILE}\PY{p}{;}
                 \PY{n}{kk} \PY{o}{\PYZlt{}} \PY{n}{size}\PY{o}{/}\PY{l+m+mi}{2L}\PY{p}{;} \PY{n}{kk}\PY{o}{+=}\PY{n}{kTILE}\PY{p}{)} \PY{p}{\PYZob{}}

  \PY{c+c1}{// Unfused loop to precompute trigonometric}
  \PY{c+c1}{// functions with vector operations}
 \PY{k+kt}{float} \PY{n}{coslist}\PY{p}{[}\PY{n}{kTILE}\PY{p}{]} \PY{n}{ALIGNED}\PY{p}{,}
       \PY{n}{sinlist}\PY{p}{[}\PY{n}{kTILE}\PY{p}{]} \PY{n}{ALIGNED}\PY{p}{,}
 \PY{k}{for} \PY{p}{(}\PY{k+kt}{int} \PY{n}{i} \PY{o}{=} \PY{l+m+mi}{0}\PY{p}{;} \PY{n}{i} \PY{o}{\PYZlt{}} \PY{n}{kTILE}\PY{p}{;} \PY{n}{i}\PY{o}{++}\PY{p}{)} \PY{p}{\PYZob{}}
  \PY{n}{coslist}\PY{p}{[}\PY{n}{i}\PY{p}{]} \PY{o}{=} \PY{n}{cosf}\PY{p}{((}\PY{n}{kk}\PY{o}{+}\PY{n}{i}\PY{p}{)}\PY{o}{*}\PY{n}{trigconst}\PY{p}{);}
  \PY{n}{sinlist}\PY{p}{[}\PY{n}{i}\PY{p}{]} \PY{o}{=} \PY{n}{sinf}\PY{p}{((}\PY{n}{kk}\PY{o}{+}\PY{n}{i}\PY{p}{)}\PY{o}{*}\PY{n}{trigconst}\PY{p}{);}
 \PY{p}{\PYZcb{}}

 \PY{c+c1}{// Iterating within the tile}
 \PY{k}{for} \PY{p}{(}\PY{k+kt}{long} \PY{n}{k} \PY{o}{=} \PY{n}{kk}\PY{p}{;} \PY{n}{k} \PY{o}{\PYZlt{}} \PY{n}{kk}\PY{o}{+}\PY{n}{kTILE}\PY{p}{;} \PY{n}{k}\PY{o}{++} \PY{p}{\PYZob{}}
  \PY{n}{target}\PY{p}{[}\PY{l+m+mi}{2}\PY{o}{*}\PY{n}{k}\PY{p}{]}\PY{o}{=}\PY{n}{evens}\PY{p}{[}\PY{l+m+mi}{2}\PY{o}{*}\PY{n}{k}\PY{p}{]}\PY{o}{+}\PY{n}{odds}\PY{p}{[}\PY{l+m+mi}{2}\PY{o}{*}\PY{n}{k}\PY{p}{]}\PY{o}{*}\PY{n}{coslist}\PY{p}{[}\PY{n}{k}\PY{p}{]}
                      \PY{o}{\PYZhy{}}\PY{n}{odds}\PY{p}{[}\PY{l+m+mi}{2}\PY{o}{*}\PY{n}{k}\PY{o}{+}\PY{l+m+mi}{1}\PY{p}{]}\PY{o}{*}\PY{n}{sinlist}\PY{p}{[}\PY{n}{k}\PY{p}{];}
  \PY{n}{target}\PY{p}{[}\PY{l+m+mi}{2}\PY{o}{*}\PY{n}{k}\PY{o}{+}\PY{l+m+mi}{1}\PY{p}{]}\PY{o}{=}\PY{n}{evens}\PY{p}{[}\PY{l+m+mi}{2}\PY{o}{*}\PY{n}{k}\PY{o}{+}\PY{l+m+mi}{1}\PY{p}{]}\PY{o}{+}
                      \PY{n}{odds}\PY{p}{[}\PY{l+m+mi}{2}\PY{o}{*}\PY{n}{k}\PY{p}{]}\PY{o}{*}\PY{n}{sinlist}\PY{p}{[}\PY{n}{k}\PY{o}{\PYZhy{}}\PY{n}{kk}\PY{p}{]}
                   \PY{o}{+}\PY{n}{odds}\PY{p}{[}\PY{l+m+mi}{2}\PY{o}{*}\PY{n}{k}\PY{o}{+}\PY{l+m+mi}{1}\PY{p}{]}\PY{o}{*}\PY{n}{coslist}\PY{p}{[}\PY{n}{k}\PY{o}{\PYZhy{}}\PY{n}{kk}\PY{p}{];}
   \PY{c+c1}{//...}
 \PY{p}{\PYZcb{}}
\PY{p}{\PYZcb{}}
\end{Verbatim}
\vskip -0.5em
\caption{Improved (still not optimal) reassembly code. Strip-mining helps to vectorize the trigonometric functions. \label{code:stripminedreassembly}}
\end{listing}

The second optimization of vectorization relies on the Intel Cilk Plus array notation to assist the vectorization of operation on strided data.
Indeed, the \texttt{for}-loop in line 24 of Listing~\ref{code:stripminedreassembly} mixes data with a stride of 1 (in \texttt{sinlist} and \texttt{coslist}) and stride 2 (\texttt{evens}, \texttt{odds} and \texttt{target}). We have found that re-writing this loop with array notation as shown in Listing~\ref{code:arraynotationreassembly} improves the efficiency of automatic vectorization. 
Array notation is not a standard part of the C++ language; it is a language extension provided by the Intel Cilk Plus framework.
It is supported by the Intel C++ compiler and by GCC \cite{gcc-cilk}.

\begin{listing}[H]
\begin{Verbatim}[commandchars=\\\{\},numbers=left,firstnumber=1,stepnumber=1, ,fontsize=\scriptsize ,frame=single,xleftmargin=10pt,numbersep=2pt]
\PY{n}{cilk\PYZus{}for}\PY{p}{(}\PY{k+kt}{long} \PY{n}{kk} \PY{o}{=} \PY{n}{kTILE}\PY{p}{;}
                 \PY{n}{kk} \PY{o}{\PYZlt{}} \PY{n}{size}\PY{o}{/}\PY{l+m+mi}{2L}\PY{p}{;} \PY{n}{kk}\PY{o}{+=}\PY{n}{kTILE}\PY{p}{)} \PY{p}{\PYZob{}}

 \PY{k+kt}{float} \PY{n}{coslist}\PY{p}{[}\PY{n}{kTILE}\PY{p}{]} \PY{n}{ALIGNED}\PY{p}{,}
       \PY{n}{sinlist}\PY{p}{[}\PY{n}{kTILE}\PY{p}{]} \PY{n}{ALIGNED}\PY{p}{,}
       \PY{n}{evenrek}\PY{p}{[}\PY{n}{kTILE}\PY{p}{]} \PY{n}{ALIGNED}\PY{p}{,}
       \PY{n}{evenimk}\PY{p}{[}\PY{n}{kTILE}\PY{p}{]} \PY{n}{ALIGNED}\PY{p}{,}
 \PY{c+c1}{// ...}

\PY{c+cp}{\PYZsh{}pragma simd}
\PY{c+cp}{\PYZsh{}pragma vector aligned}
  \PY{k}{for} \PY{p}{(}\PY{k+kt}{int} \PY{n}{i} \PY{o}{=} \PY{l+m+mi}{0}\PY{p}{;} \PY{n}{i} \PY{o}{\PYZlt{}} \PY{n}{kTILE}\PY{p}{;} \PY{n}{i}\PY{o}{++}\PY{p}{)} \PY{p}{\PYZob{}}
    \PY{n}{sinlist}\PY{p}{[}\PY{n}{i}\PY{p}{]} \PY{o}{=} \PY{n}{sinf}\PY{p}{(}\PY{n}{theta}\PY{o}{+}\PY{n}{i}\PY{o}{*}\PY{n}{trigconst}\PY{p}{);}
    \PY{n}{coslist}\PY{p}{[}\PY{n}{i}\PY{p}{]} \PY{o}{=} \PY{n}{cosf}\PY{p}{(}\PY{n}{theta}\PY{o}{+}\PY{n}{i}\PY{o}{*}\PY{n}{trigconst}\PY{p}{);}
  \PY{p}{\PYZcb{}}

  \PY{c+c1}{// Iterating within the tile (array notation)}

  \PY{c+c1}{// Gather elements with a stride of 2}
  \PY{n}{evenrek}\PY{p}{[}\PY{o}{:}\PY{p}{]} \PY{o}{=} \PY{n}{evens}\PY{p}{[}\PY{n}{kk}  \PY{o}{:}\PY{n}{kTILE}\PY{o}{:}\PY{l+m+mi}{2}\PY{p}{];}
  \PY{n}{evenimk}\PY{p}{[}\PY{o}{:}\PY{p}{]} \PY{o}{=} \PY{n}{evens}\PY{p}{[}\PY{n}{kk}\PY{o}{+}\PY{l+m+mi}{1}\PY{o}{:}\PY{n}{kTILE}\PY{o}{:}\PY{l+m+mi}{2}\PY{p}{];}
  \PY{n}{oddrek} \PY{p}{[}\PY{o}{:}\PY{p}{]} \PY{o}{=} \PY{n}{odds} \PY{p}{[}\PY{n}{kk}  \PY{o}{:}\PY{n}{kTILE}\PY{o}{:}\PY{l+m+mi}{2}\PY{p}{];}
  \PY{n}{oddimk} \PY{p}{[}\PY{o}{:}\PY{p}{]} \PY{o}{=} \PY{n}{odds} \PY{p}{[}\PY{n}{kk}\PY{o}{+}\PY{l+m+mi}{1}\PY{o}{:}\PY{n}{kTILE}\PY{o}{:}\PY{l+m+mi}{2}\PY{p}{];}

  \PY{c+c1}{// Reassembl \PYZam{} scatter into stride\PYZhy{}2 array}
  \PY{n}{target}\PY{p}{[}\PY{n}{kk}  \PY{o}{:}\PY{n}{kTILE}\PY{o}{:}\PY{l+m+mi}{2}\PY{p}{]} \PY{o}{=} \PY{n}{evenrek}\PY{p}{[}\PY{o}{:}\PY{p}{]} \PY{o}{+}
   \PY{n}{coslist}\PY{p}{[}\PY{o}{:}\PY{p}{]}\PY{o}{*}\PY{n}{oddrek}\PY{p}{[}\PY{o}{:}\PY{p}{]}\PY{o}{\PYZhy{}}\PY{n}{sinlist}\PY{p}{[}\PY{o}{:}\PY{p}{]}\PY{o}{*}\PY{n}{oddimk}\PY{p}{[}\PY{o}{:}\PY{p}{];}
  \PY{n}{target}\PY{p}{[}\PY{n}{kk}\PY{o}{+}\PY{l+m+mi}{1}\PY{o}{:}\PY{n}{kTILE}\PY{o}{:}\PY{l+m+mi}{2}\PY{p}{]} \PY{o}{=} \PY{n}{evenimk}\PY{p}{[}\PY{o}{:}\PY{p}{]} \PY{o}{+}
   \PY{n}{sinlist}\PY{p}{[}\PY{o}{:}\PY{p}{]}\PY{o}{*}\PY{n}{oddrek}\PY{p}{[}\PY{o}{:}\PY{p}{]}\PY{o}{+}\PY{n}{coslist}\PY{p}{[}\PY{o}{:}\PY{p}{]}\PY{o}{*}\PY{n}{oddimk}\PY{p}{[}\PY{o}{:}\PY{p}{];}

   \PY{c+c1}{//...}
\PY{p}{\PYZcb{}}
\end{Verbatim}
\vskip -0.5em
\caption{Further optimized (still not optimal) reassembly code. Array notation helps the compiler to vectorize operations with strided data accesses. \label{code:arraynotationreassembly}}
\end{listing}

\begin{figure}[H]
\begin{center}
\includegraphics[width=0.5\textwidth]{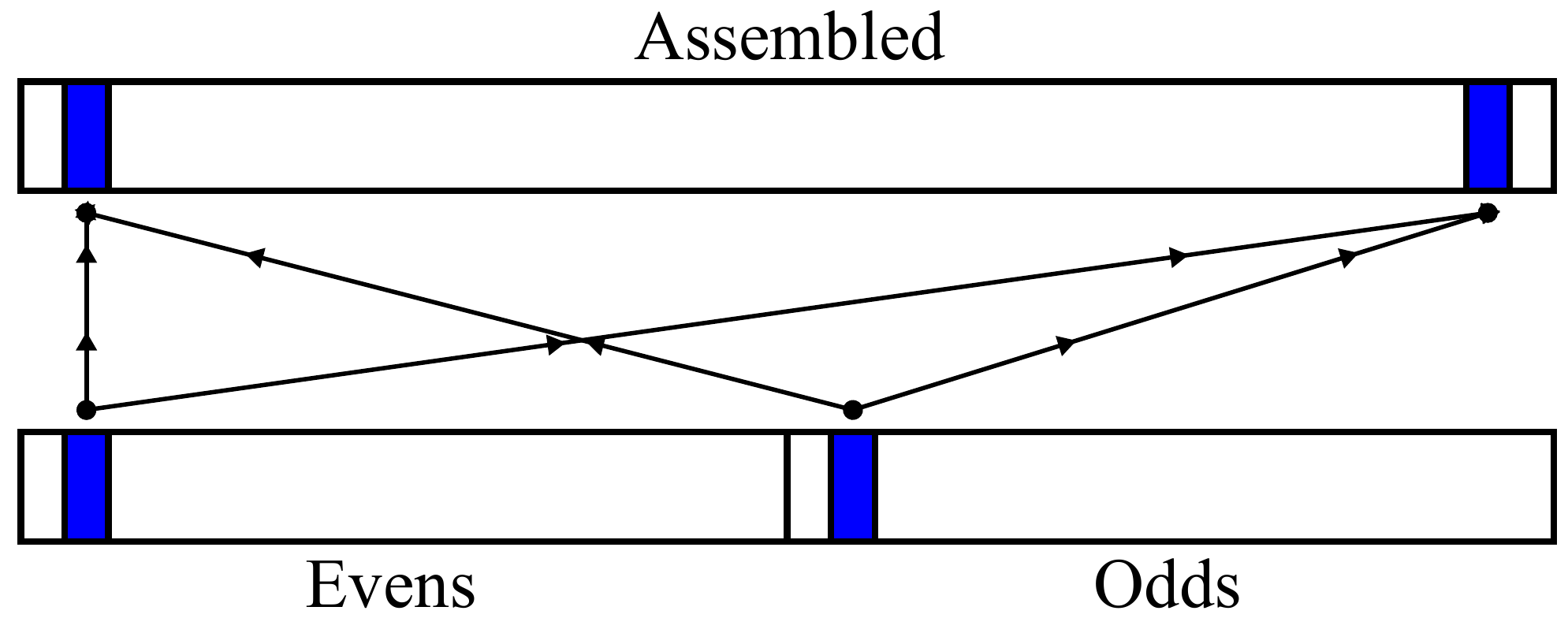}
\caption{Diagram showing the memory locations in a CT assembly of an element. Because of the location of the elements, this algorithm can not be done in-place.\label{diag:CTAssembly}}
\end{center}
\end{figure}

Finally, one more target of optimization in the reassembly code is memory access. The unoptimized code in Listing~\ref{code:arraynotationreassembly} combines two elements from input arrays and writes out the result into a separate output array. This means that in this code, reassembly is done {\em out-of-place}. Indeed, the array index of the destination element $F_k$ is not the same as the index of the source elements $E_k$ and $O_k$.
Diagram (\ref{diag:CTAssembly}) demonstrates this problem.

\begin{figure}[H]
\begin{center}
\includegraphics[width=0.5\textwidth]{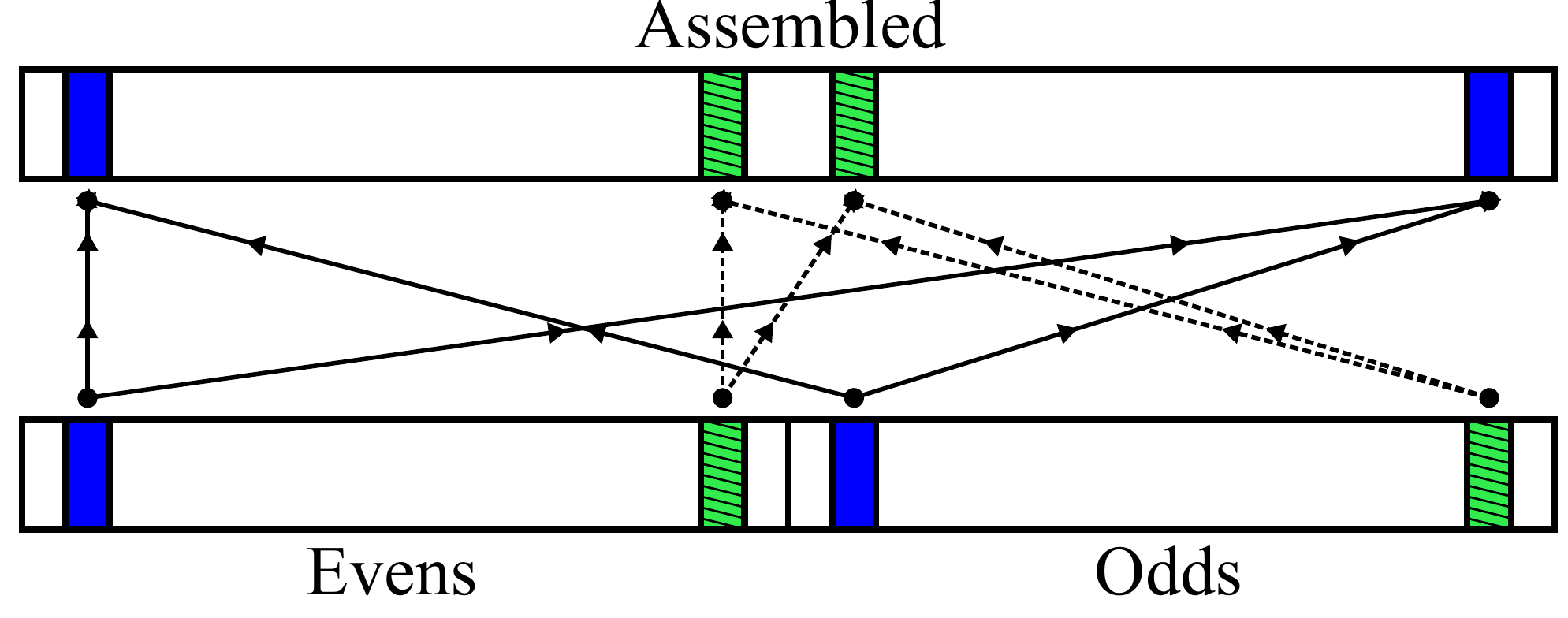}
\caption{Diagram showing the memory locations in a CT assembly in pairs. With this method the memory locations overlap, and in-place CT Assembly is possible.\label{diag:CTAssemblyPair}}
\end{center}
\end{figure}

However, this difficulty can be overcome by performing reassembly in pairs of elements.
Due to the symmetry of the operation, by combining two different steps in reassembly, it is possible to get an assembly step where the memory location for the input is the same as the destination memory location for the output.
Indeed, upon the assembly of $E_k$ and $O_k$, we have to write $F_k$ and $F_{N-k-1}$. If at the same time we assemble $E_{N/2-k-1}$ and $O_{N/2-k-1}$, then we can write $F_{N/2-k-1}$ and $F_{N-(N/2-k-1)-1} = F_{N/2+k}$. The offset of the latter element coincides with the offset of $O_k$. Therefore, it is possible to read the values of $E$ and $O$ elements and write the values of the corresponding $F$ elements into the same memory locations.
Diagram (\ref{diag:CTAssemblyPair}) shows this assembly-in-pair scheme.
Using this method, in the optimized version of element combination function (Listing~\ref{lst:reassembly-opt}) we perform reassembly {\em in-place}.
By doing this, we reduce the number of memory accesses by a factor of 2.

\begin{listing}[H]
\begin{Verbatim}[commandchars=\\\{\},numbers=left,firstnumber=1,stepnumber=1, ,fontsize=\scriptsize ,frame=single,xleftmargin=10pt,numbersep=2pt]
\PY{n}{cilk\PYZus{}for} \PY{p}{(}\PY{k+kt}{long} \PY{n}{kk} \PY{o}{=} \PY{n}{kTILE}\PY{p}{;} \PY{n}{kk} \PY{o}{\PYZlt{}} \PY{n}{halfsize}\PY{o}{/}\PY{l+m+mi}{2L}\PY{p}{;}
                              \PY{n}{kk} \PY{o}{+=} \PY{n}{kTILE}\PY{p}{)} \PY{p}{\PYZob{}}
  \PY{k+kt}{float} \PY{n}{coslist}\PY{p}{[}\PY{n}{kTILE}\PY{p}{]} \PY{n}{ALIGNED}\PY{p}{,}
        \PY{n}{sinlist}\PY{p}{[}\PY{n}{kTILE}\PY{p}{]} \PY{n}{ALIGNED}\PY{p}{,}
        \PY{n}{evenrek}\PY{p}{[}\PY{n}{kTILE}\PY{p}{]} \PY{n}{ALIGNED}\PY{p}{,}
        \PY{n}{evenimk}\PY{p}{[}\PY{n}{kTILE}\PY{p}{]} \PY{n}{ALIGNED}\PY{p}{,}
        \PY{c+c1}{//...}

  \PY{c+c1}{// Vectorized twiddle factor precomputation}
  \PY{k}{const} \PY{k+kt}{float} \PY{n}{theta}\PY{o}{=}\PY{p}{(}\PY{k+kt}{float}\PY{p}{)(}\PY{n}{kk}\PY{o}{/}\PY{l+m+mi}{2}\PY{p}{)}\PY{o}{*}\PY{n}{trigconst}\PY{p}{;}
\PY{c+cp}{\PYZsh{}pragma simd}
\PY{c+cp}{\PYZsh{}pragma vector aligned}
  \PY{k}{for} \PY{p}{(}\PY{k+kt}{int} \PY{n}{i} \PY{o}{=} \PY{l+m+mi}{0}\PY{p}{;} \PY{n}{i} \PY{o}{\PYZlt{}} \PY{n}{kTILE}\PY{p}{;} \PY{n}{i}\PY{o}{++}\PY{p}{)} \PY{p}{\PYZob{}}
    \PY{n}{sinlist}\PY{p}{[}\PY{n}{i}\PY{p}{]} \PY{o}{=} \PY{n}{sinf}\PY{p}{(}\PY{n}{theta}\PY{o}{+}\PY{n}{i}\PY{o}{*}\PY{n}{trigconst}\PY{p}{);}
    \PY{n}{coslist}\PY{p}{[}\PY{n}{i}\PY{p}{]} \PY{o}{=} \PY{n}{cosf}\PY{p}{(}\PY{n}{theta}\PY{o}{+}\PY{n}{i}\PY{o}{*}\PY{n}{trigconst}\PY{p}{);}
  \PY{p}{\PYZcb{}}

  \PY{c+c1}{// Gather stride\PYZhy{}2 data for first pair}
  \PY{n}{evenrek}\PY{p}{[}\PY{o}{:}\PY{p}{]} \PY{o}{=} \PY{n}{evens}\PY{p}{[}\PY{n}{kk}  \PY{o}{:}\PY{n}{kTILE}\PY{o}{:}\PY{l+m+mi}{2}\PY{p}{];}
  \PY{n}{evenimk}\PY{p}{[}\PY{o}{:}\PY{p}{]} \PY{o}{=} \PY{n}{evens}\PY{p}{[}\PY{n}{kk}\PY{o}{+}\PY{l+m+mi}{1}\PY{o}{:}\PY{n}{kTILE}\PY{o}{:}\PY{l+m+mi}{2}\PY{p}{];}
  \PY{n}{oddrek} \PY{p}{[}\PY{o}{:}\PY{p}{]} \PY{o}{=} \PY{n}{odds} \PY{p}{[}\PY{n}{kk}  \PY{o}{:}\PY{n}{kTILE}\PY{o}{:}\PY{l+m+mi}{2}\PY{p}{];}
  \PY{n}{oddimk} \PY{p}{[}\PY{o}{:}\PY{p}{]} \PY{o}{=} \PY{n}{odds} \PY{p}{[}\PY{n}{kk}\PY{o}{+}\PY{l+m+mi}{1}\PY{o}{:}\PY{n}{kTILE}\PY{o}{:}\PY{l+m+mi}{2}\PY{p}{];}

  \PY{c+c1}{// Reassemble first pair}
  \PY{c+c1}{// and scatter results into stride\PYZhy{}2 array}
  \PY{n}{evens}\PY{p}{[}\PY{n}{kk}  \PY{o}{:}\PY{n}{kTILE}\PY{o}{:}\PY{l+m+mi}{2}\PY{p}{]} \PY{o}{=} \PY{n}{evenrek}\PY{p}{[}\PY{o}{:}\PY{p}{]} \PY{o}{+}
   \PY{n}{coslist}\PY{p}{[}\PY{o}{:}\PY{p}{]}\PY{o}{*}\PY{n}{oddrek}\PY{p}{[}\PY{o}{:}\PY{p}{]}\PY{o}{\PYZhy{}}\PY{n}{sinlist}\PY{p}{[}\PY{o}{:}\PY{p}{]}\PY{o}{*}\PY{n}{oddimk}\PY{p}{[}\PY{o}{:}\PY{p}{];}
  \PY{n}{evens}\PY{p}{[}\PY{n}{kk}\PY{o}{+}\PY{l+m+mi}{1}\PY{o}{:}\PY{n}{kTILE}\PY{o}{:}\PY{l+m+mi}{2}\PY{p}{]} \PY{o}{=} \PY{n}{evenimk}\PY{p}{[}\PY{o}{:}\PY{p}{]} \PY{o}{+}
   \PY{n}{sinlist}\PY{p}{[}\PY{o}{:}\PY{p}{]}\PY{o}{*}\PY{n}{oddrek}\PY{p}{[}\PY{o}{:}\PY{p}{]}\PY{o}{+}\PY{n}{coslist}\PY{p}{[}\PY{o}{:}\PY{p}{]}\PY{o}{*}\PY{n}{oddimk}\PY{p}{[}\PY{o}{:}\PY{p}{];}

  \PY{c+c1}{// Start gather second pair (mirror)}
  \PY{n}{oddmirrek}\PY{p}{[}\PY{o}{:}\PY{p}{]} \PY{o}{=} \PY{n}{odds}\PY{p}{[}\PY{n}{size}\PY{o}{\PYZhy{}}\PY{n}{kk}  \PY{o}{:}\PY{n}{kTILE}\PY{o}{:\PYZhy{}}\PY{l+m+mi}{2}\PY{p}{];}
  \PY{n}{oddmirimk}\PY{p}{[}\PY{o}{:}\PY{p}{]} \PY{o}{=} \PY{n}{odds}\PY{p}{[}\PY{n}{size}\PY{o}{\PYZhy{}}\PY{n}{kk}\PY{o}{+}\PY{l+m+mi}{1}\PY{o}{:}\PY{n}{kTILE}\PY{o}{:\PYZhy{}}\PY{l+m+mi}{2}\PY{p}{];}

  \PY{c+c1}{// Finish reassembly of first pair}
  \PY{n}{odds}\PY{p}{[}\PY{n}{size}\PY{o}{\PYZhy{}}\PY{n}{kk}  \PY{o}{:}\PY{n}{kTILE}\PY{o}{:\PYZhy{}}\PY{l+m+mi}{2}\PY{p}{]} \PY{o}{=}  \PY{n}{evenrek}\PY{p}{[}\PY{o}{:}\PY{p}{]} \PY{o}{\PYZhy{}}
   \PY{n}{coslist}\PY{p}{[}\PY{o}{:}\PY{p}{]}\PY{o}{*}\PY{n}{oddrek}\PY{p}{[}\PY{o}{:}\PY{p}{]}\PY{o}{+}\PY{n}{sinlist}\PY{p}{[}\PY{o}{:}\PY{p}{]}\PY{o}{*}\PY{n}{oddimk}\PY{p}{[}\PY{o}{:}\PY{p}{];}
  \PY{n}{odds}\PY{p}{[}\PY{n}{size}\PY{o}{\PYZhy{}}\PY{n}{kk}\PY{o}{+}\PY{l+m+mi}{1}\PY{o}{:}\PY{n}{kTILE}\PY{o}{:\PYZhy{}}\PY{l+m+mi}{2}\PY{p}{]} \PY{o}{=} \PY{o}{\PYZhy{}}\PY{n}{evenimk}\PY{p}{[}\PY{o}{:}\PY{p}{]} \PY{o}{+}
   \PY{n}{sinlist}\PY{p}{[}\PY{o}{:}\PY{p}{]}\PY{o}{*}\PY{n}{oddrek}\PY{p}{[}\PY{o}{:}\PY{p}{]}\PY{o}{+}\PY{n}{coslist}\PY{p}{[}\PY{o}{:}\PY{p}{]}\PY{o}{*}\PY{n}{oddimk}\PY{p}{[}\PY{o}{:}\PY{p}{];}

  \PY{c+c1}{// Finish gathering second pair (mirror)}
  \PY{n}{evenmirrek}\PY{p}{[}\PY{o}{:}\PY{p}{]} \PY{o}{=} \PY{n}{evens}\PY{p}{[}\PY{n}{size}\PY{o}{\PYZhy{}}\PY{n}{kk}  \PY{o}{:}\PY{n}{kTILE}\PY{o}{:\PYZhy{}}\PY{l+m+mi}{2}\PY{p}{];}
  \PY{n}{evenmirimk}\PY{p}{[}\PY{o}{:}\PY{p}{]} \PY{o}{=} \PY{n}{evens}\PY{p}{[}\PY{n}{size}\PY{o}{\PYZhy{}}\PY{n}{kk}\PY{o}{+}\PY{l+m+mi}{1}\PY{o}{:}\PY{n}{kTILE}\PY{o}{:\PYZhy{}}\PY{l+m+mi}{2}\PY{p}{];}

  \PY{c+c1}{// Reassemble second pair of elements}
  \PY{c+c1}{// and scatter results into stride\PYZhy{}2 array}
  \PY{n}{evens}\PY{p}{[}\PY{n}{size}\PY{o}{\PYZhy{}}\PY{n}{kk}  \PY{o}{:}\PY{n}{kTILE}\PY{o}{:\PYZhy{}}\PY{l+m+mi}{2}\PY{p}{]}\PY{o}{=} \PY{n}{evenmirrek}\PY{p}{[}\PY{o}{:}\PY{p}{]} \PY{o}{\PYZhy{}}
                   \PY{n}{sinlist}\PY{p}{[}\PY{o}{:}\PY{p}{]}\PY{o}{*}\PY{n}{oddmirrek}\PY{p}{[}\PY{o}{:}\PY{p}{]}
                  \PY{o}{+}\PY{n}{coslist}\PY{p}{[}\PY{o}{:}\PY{p}{]}\PY{o}{*}\PY{n}{oddmirimk}\PY{p}{[}\PY{o}{:}\PY{p}{];}
  \PY{n}{evens}\PY{p}{[}\PY{n}{size}\PY{o}{\PYZhy{}}\PY{n}{kk}\PY{o}{+}\PY{l+m+mi}{1}\PY{o}{:}\PY{n}{kTILE}\PY{o}{:\PYZhy{}}\PY{l+m+mi}{2}\PY{p}{]}\PY{o}{=} \PY{n}{evenmirimk}\PY{p}{[}\PY{o}{:}\PY{p}{]} \PY{o}{\PYZhy{}}
                   \PY{n}{coslist}\PY{p}{[}\PY{o}{:}\PY{p}{]}\PY{o}{*}\PY{n}{oddmirrek}\PY{p}{[}\PY{o}{:}\PY{p}{]}
                  \PY{o}{\PYZhy{}}\PY{n}{sinlist}\PY{p}{[}\PY{o}{:}\PY{p}{]}\PY{o}{*}\PY{n}{oddmirimk}\PY{p}{[}\PY{o}{:}\PY{p}{];}
  \PY{n}{odds} \PY{p}{[}\PY{n}{kk}       \PY{o}{:}\PY{n}{kTILE}\PY{o}{:} \PY{l+m+mi}{2}\PY{p}{]}\PY{o}{=} \PY{n}{evenmirrek}\PY{p}{[}\PY{o}{:}\PY{p}{]} \PY{o}{+}
                   \PY{n}{sinlist}\PY{p}{[}\PY{o}{:}\PY{p}{]}\PY{o}{*}\PY{n}{oddmirrek}\PY{p}{[}\PY{o}{:}\PY{p}{]}
                  \PY{o}{\PYZhy{}}\PY{n}{coslist}\PY{p}{[}\PY{o}{:}\PY{p}{]}\PY{o}{*}\PY{n}{oddmirimk}\PY{p}{[}\PY{o}{:}\PY{p}{];}
  \PY{n}{odds} \PY{p}{[}\PY{n}{kk}\PY{o}{+}\PY{l+m+mi}{1}     \PY{o}{:}\PY{n}{kTILE}\PY{o}{:} \PY{l+m+mi}{2}\PY{p}{]}\PY{o}{=\PYZhy{}}\PY{n}{evenmirimk}\PY{p}{[}\PY{o}{:}\PY{p}{]} \PY{o}{\PYZhy{}}
                   \PY{n}{coslist}\PY{p}{[}\PY{o}{:}\PY{p}{]}\PY{o}{*}\PY{n}{oddmirrek}\PY{p}{[}\PY{o}{:}\PY{p}{]}
                   \PY{o}{\PYZhy{}}\PY{n}{sinlist}\PY{p}{[}\PY{o}{:}\PY{p}{]}\PY{o}{*}\PY{n}{oddmirimk}\PY{p}{[}\PY{o}{:}\PY{p}{];}
\PY{p}{\PYZcb{}}
\end{Verbatim}
\vskip -0.5em
\caption{Optimized main loop in the reassembly function.\label{lst:reassembly-opt}}
\end{listing}

The code shown in Listing~\ref{lst:reassembly-opt} is the final version that we adopted in EFFT. It is optimized with the following methods:
\begin{enumerate}
\tighterenum
\item Strip-mining to precompute the trigonometric functions with vector operations,
\item Array notation to help the compiler vectorize loops with non-unit stride access, and
\item Performing reassembly on symmetrically located pairs of elements in each iteration in order to perform the job {\em in-place}, reducing the number of memory accesses.
\end{enumerate}

Listing~\ref{lst:reassembly-opt} does not contain the code for the processing of elements $k=0$ and $k=N/2$, because they are stored in a different way from the rest of the elements (see Equation~(\ref{eq:dataformat}). It also does not contain the code for processing of elements $k=1$ through $k=$~\texttt{kTILE}$-1$. These elements are processed separately because they do not comprise a full ``tile'', and processing them together with the rest of the tiles would require protection with branches, which would ruin vectorization performance. For the complete code listing, refer to the source code available at \cite{efft}.

\newpage
\section{Using the EFFT Library}\label{sec:EFFTdoc}

EFFT is a C++ library that implements \texttt{class EFFT\_Transform}:

\begin{listing}[H]
\begin{Verbatim}[commandchars=\\\{\},numbers=left,firstnumber=1,stepnumber=1, ,fontsize=\scriptsize ,frame=single,xleftmargin=10pt,numbersep=2pt]
\PY{k}{class} \PY{n+nc}{EFFT\PYZus{}Transform} \PY{p}{\PYZob{}}

\PY{c+c1}{// Private members not shown}

\PY{n+nl}{public:}
  \PY{n}{EFFT\PYZus{}Transform}\PY{p}{(} \PY{c+c1}{// Constructor}
        \PY{k}{const} \PY{k+kt}{long} \PY{n}{n}\PY{p}{,} \PY{k}{const} \PY{k+kt}{int} \PY{n}{splits}\PY{p}{);}
  \PY{o}{\PYZti{}}\PY{n}{EFFT\PYZus{}Transform}\PY{p}{();} \PY{c+c1}{// Destructor}

  \PY{k+kt}{void} \PY{n+nf}{RunTransform}\PY{p}{();} \PY{c+c1}{// Main method}
  \PY{k+kt}{float}\PY{o}{*} \PY{n+nf}{Data}\PY{p}{();} \PY{c+c1}{// Accessor to input}
  \PY{k+kt}{float}\PY{o}{*} \PY{n+nf}{Result}\PY{p}{();} \PY{c+c1}{// Accessor to output}
\PY{p}{\PYZcb{};}
\end{Verbatim}
\vskip -0.5em
\caption{Declaration of \texttt{EFFT\_Transform}\label{code:efftheader}}
\end{listing}

In order to perform a DFT, an object of type \texttt{class EFFT\_Transform} must be initialized. Its constructor requires two inputs: size ($N$) and splits ($s$).
The size argument is the size of the full transform of type \texttt{long}.
Splits is the number of Radix-2 CT that EFFT applies on the data set.
Note that our current implementation of EFFT only accepts sizes that are multiples of $2^{(s+8)}$.
This restriction is the consequence of the Radix-2 CT algorithm, as well as the tiling optimization we have implemented.

The intialization of an \texttt{EFFT\_Transform} object may take some time, especially for large arrays.
However, once it has been created, it can be reused as many times as the user needs.
As discussed in Section (\ref{sec:EFFTlibrary}), the \texttt{EFFT\_Transform} internally allocates the memory space it needs, including the data array.

After initialization of the main class, the user populates the data array using the accessor \texttt{EFFT\_Transform::Data()}.
Note that this is a pointer to an array of type \texttt{float}; EFFT currently does not support double precision.

After the array was populated with data, the user should simply call the method \texttt{EFFT\_Transform::RunTransform()} to carry out the DFT on the data array.
The pointer to the result of the DFT can be obtained with the \texttt{EFFT\_Transform::Result()} accessor method.
The current implementation of EFFT does the Transform out-of-place, so the data array will retain the input and \texttt{EFFT\_Transform::Result()} and \texttt{EFFT\_Transform::Data()} will return different pointers.

Listing (\ref{code:usingEFFT}) shows a sample implementation of a 1D DFT using the EFFT library. 

\begin{listing}[H]
\begin{Verbatim}[commandchars=\\\{\},numbers=left,firstnumber=1,stepnumber=1, ,fontsize=\scriptsize ,frame=single,xleftmargin=10pt,numbersep=2pt]
\PY{n}{EFFT\PYZus{}Transform} \PY{n+nf}{myTransform}\PY{p}{(}\PY{n}{n}\PY{p}{,} \PY{n}{numofsplits}\PY{p}{);}

\PY{c+c1}{//Getting the pointer to the data}
\PY{k+kt}{float}\PY{o}{*} \PY{n}{A} \PY{o}{=} \PY{n}{myTransform}\PY{p}{.}\PY{n}{Data}\PY{p}{();}

\PY{c+c1}{//Populating with random data}
\PY{n}{srand}\PY{p}{(}\PY{l+m+mi}{0}\PY{p}{);}
\PY{k}{for} \PY{p}{(}\PY{k+kt}{long} \PY{k+kt}{int} \PY{n}{i} \PY{o}{=} \PY{l+m+mi}{0}\PY{p}{;} \PY{n}{i} \PY{o}{\PYZlt{}} \PY{n}{n}\PY{p}{;} \PY{n}{i}\PY{o}{++}\PY{p}{)} \PY{p}{\PYZob{}}
  \PY{n}{A}\PY{p}{[}\PY{n}{i}\PY{p}{]} \PY{o}{=} \PY{p}{((}\PY{k+kt}{float}\PY{p}{)}\PY{n}{rand}\PY{p}{()}\PY{o}{/}\PY{p}{(}\PY{k+kt}{float}\PY{p}{)}\PY{n}{RAND\PYZus{}MAX}\PY{p}{)}\PY{o}{\PYZhy{}}\PY{l+m+mf}{0.5f}\PY{p}{;}
\PY{p}{\PYZcb{}}

\PY{c+c1}{//Applies FFT to the data}
\PY{n}{myTransform}\PY{p}{.}\PY{n}{RunTransform}\PY{p}{();}

\PY{c+c1}{//Getting the pointer to the result}
\PY{k+kt}{float}\PY{o}{*} \PY{n}{B} \PY{o}{=} \PY{n}{myTransform}\PY{p}{.}\PY{n}{Result}\PY{p}{();}
\end{Verbatim}
\vskip -0.5em
\caption{Using the EFFT library.\label{code:usingEFFT}}
\end{listing}

Number of splits, $s$, is one of the two tuning parameters available in EFFT.
Generally, the good performance is achieved when $2^{s}$ is nearest to the number of total available threads.
The other tuning parameter, number of workers, is set externally by either setting the environment variable \texttt{CILK\_NWORKERS} or invoking \texttt{\_\_cilkrts\_set\_param("nworkers","Wkrs")} where \texttt{Wkrs} is the number of workers to use. 
For example, in order to use 8 workers, either set \texttt{export CILK\_NWORKERS=8} in the terminal or invoke \texttt{\_\_cilkrts\_set\_param("nworkers","8")} in the C++ code.
Generally, the good performance is achieved when the number of workers is equal to the physical core limit.
The optimal value for the two tuning parameter depends on variety of factors, such as the model of the CPU, the performance of the memory subsystem, and the DFT size.
The process of this one-time optimization is discussed further in Section~(\ref{sec:Benchmarks-tuning}).

\newpage
\section{Benchmarks}\label{sec:Benchmarks}

\subsection{System Configuration}\label{sec:Benchmarks-sysconfig}

All of the benchmarks presented in this section were taken on a \href{http://www.colfax-intl.com/nd/workstations/sxp8600.aspx}{Colfax ProEdge\TM\ SXP8600} workstation based on two-way Intel Xeon E5-2697 v2 processor (12 cores per socket, 24 cores total).
We used the Intel C++ compiler version 15.0.0.90 and Intel MKL version 11.2 on a CentOS~6.5 Linux OS. For comparing with FFTW, we used FFTW version 3.3.4 compiled with the Intel C compiler with configuration arguments \texttt{--enable-avx}, \texttt{--enable-single} and \texttt{--enable-openmp}.

\subsection{Parameter Tuning}\label{sec:Benchmarks-tuning}

As discussed in Section (\ref{sec:EFFTdoc}), the tuning of EFFT is done by scanning the 2D parameter space of the number of splits, $s$, and the number of Intel Cilk Plus workers.
Fortunately, this optimization scan is a one-time requirement; the optimal values for the tuning parameters will not change for a given DFT size and hardware.
Additionally, the optimization scan is not a requirement to use the EFFT library; it is an option for users who needs the best possible performance.
EFFT package comes with a simple benchmarking code which can be used to scan the parameter space. 

\begin{figure}[H]
\begin{center}
\includegraphics[page=1,width=0.5\textwidth]{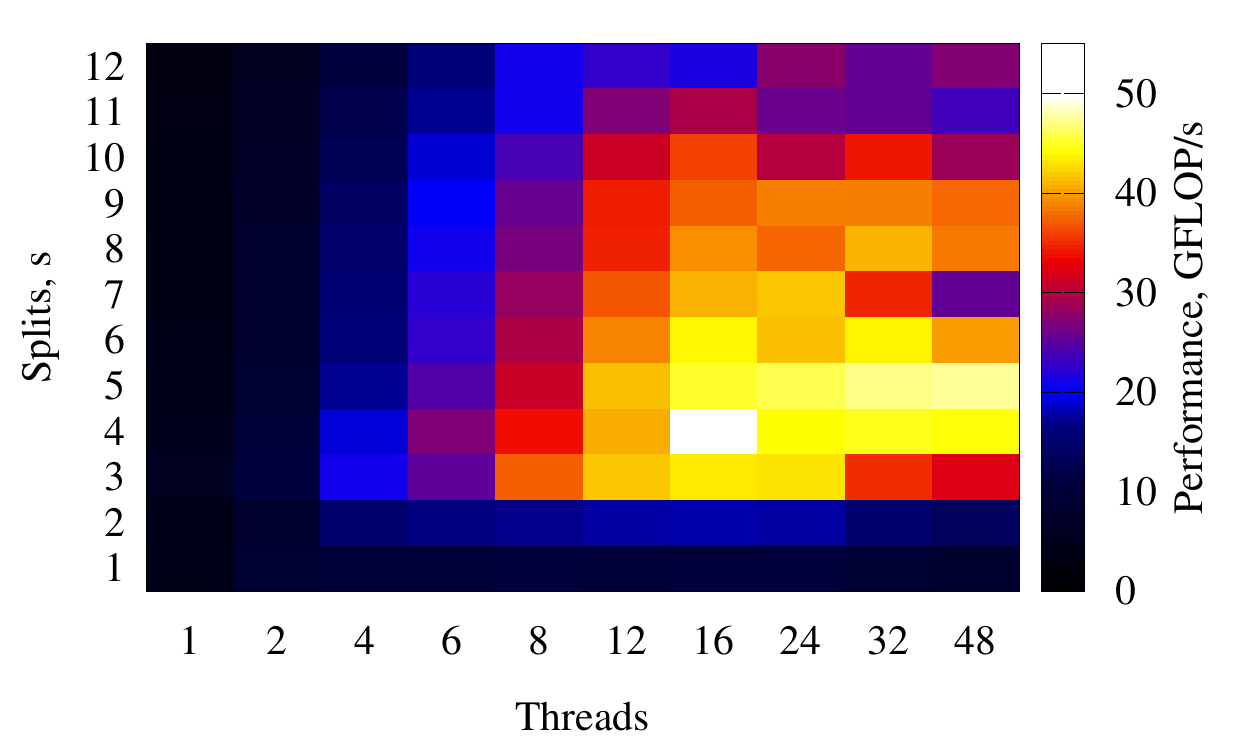}
\caption{Heat map showing the optimal number of workers and splits vs size. The areas that are ``Hot'' have better performance.\label{fig:heatmap}}
\end{center}
\end{figure}

Figure (\ref{fig:heatmap}) shows an example of this parameter space scan. Color represents the performance, in GFLOP/s, for each point in the parameter space for a DFT of size $N=2^{30}$. In this case, the optimal number of splits is $s=4$ (corresponding to $b=2^s=16$ bins), and the optimal number of workers is 16.
Here and elsewhere, performance is calculated from the wall clock run time of method \texttt{EFFT\_Transform::RunTransform()} using Equation~(\ref{eqflops}).

\begin{figure}[H]
\begin{center}
\includegraphics[page=1,width=0.5\textwidth]{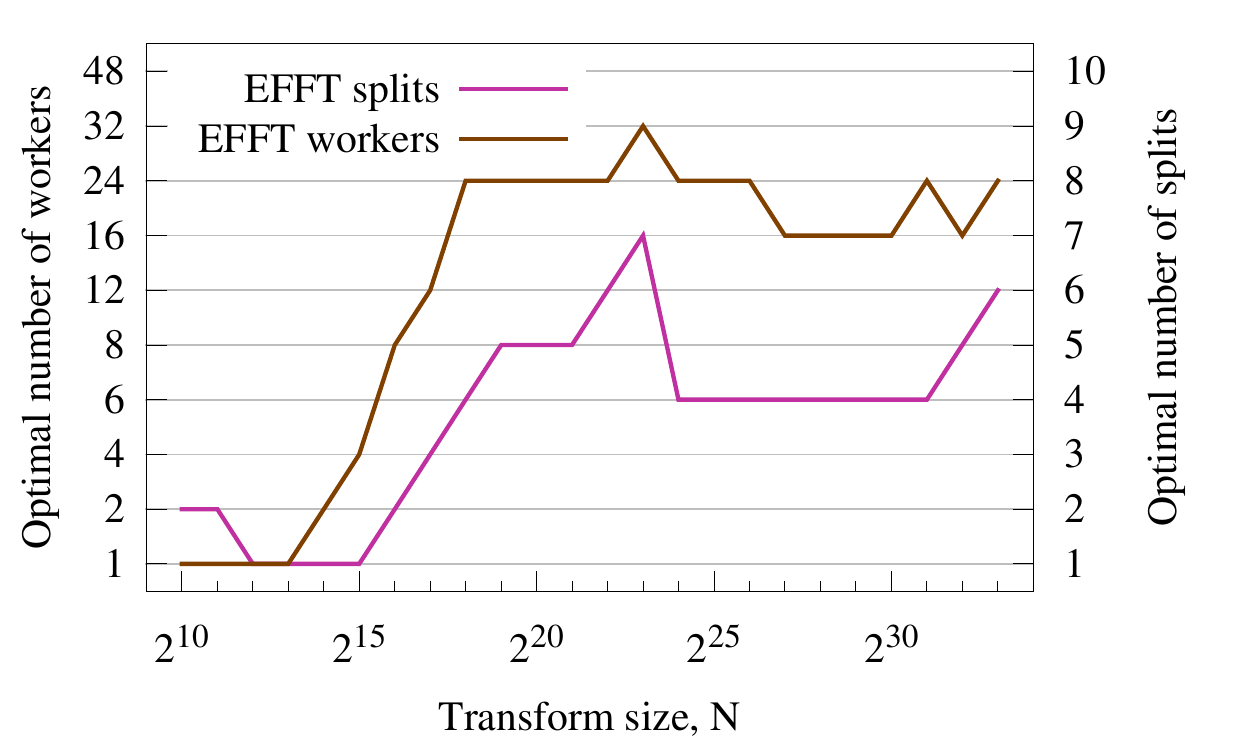}
\caption{The optimal number of workers and splits vs size. They are different for each size, and thus the parameter scan must be performed for every size to achieve the optimal performance.\label{fig:optimalthreadsspplits}}
\end{center}
\end{figure}

Figure (\ref{fig:optimalthreadsspplits}) shows the optimal number of workers and splits for EFFT on our system as a function of array size.
We use the data shown in this figure to obtain tuned performance measurements in Section~\ref{sec:Benchmarks-performance}.

In Section~\ref{sec:Benchmarks-performance}, in addition to EFFT results, we show results of multi-threaded MKL DFT implementation and of the multi-threaded FFTW library. We tuned both of these libraries by tuning the number of threads and thread affinity for each array size. In most cases, 16 or 24 threads and \texttt{KMP\_AFFINITY=scatter} yielded the best performance. For FFTW, we planned the transforms in the \texttt{FFTW\_MEASURE} mode and subsequently re-used the ``wisdom'' generated in these measurements (see \cite{fftw} for more information on FFTW planning and wisdom).

\newpage
\subsection{Performance, Memory Usage and Accuracy}\label{sec:Benchmarks-performance}

\begin{figure}[H]
\begin{center}
\includegraphics[width=0.5\textwidth]{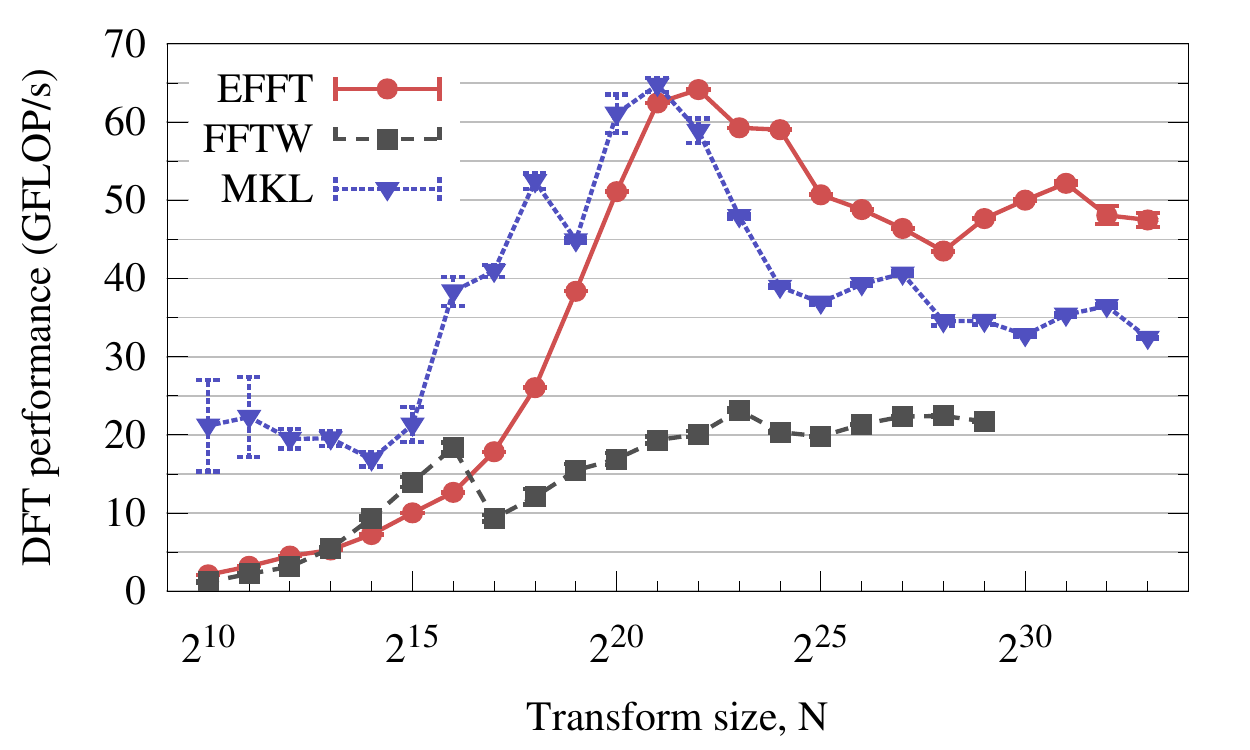}
\caption{Comparison of performance of our implementation (EFFT), Intel MKL and FFTW, as a function of array size, for $N=2^q$. The number of threads, $T$, for MKL is set to the optimal value for each point ($T=16$ for most points), and for FFTW, $T=24$ at all points. EFFT performance is reported with the optimal values of tuning parameters $T$ and $s$ at each point.\label{fig:perfsize}}
\end{center}
\end{figure}

Figure (\ref{fig:perfsize}) shows the tuned performance, in GFLOP/s, of EFFT (this work), MKL DFT and FFTW as a function of the array size. The numbers of threads for MKL and FFTW are fixed at, respectively, 16 and 24, which corresponds to values yielding the best results for most array sizes. For EFFT, the number of threads and splits vary from point to point in this plot, and are always set to the optimum value. For FFTW, we did not measure performance for arrays greater than $N=2^{30}$ because the planning time in the \texttt{FFTW\_MEASURE} mode is very long (hours to days).

According to our measurements (Figure~\ref{fig:perfsize}), EFFT outperforms FFTW for all sizes by more than 2x. It also performs better than MKL for array sizes $N>2^{22}$, achieving around performance between 45 and 55 GFLOP/s, which is 1.1x to 1.5x faster than MKL.

\begin{figure}[H]
\begin{center}
\includegraphics[width=0.5\textwidth]{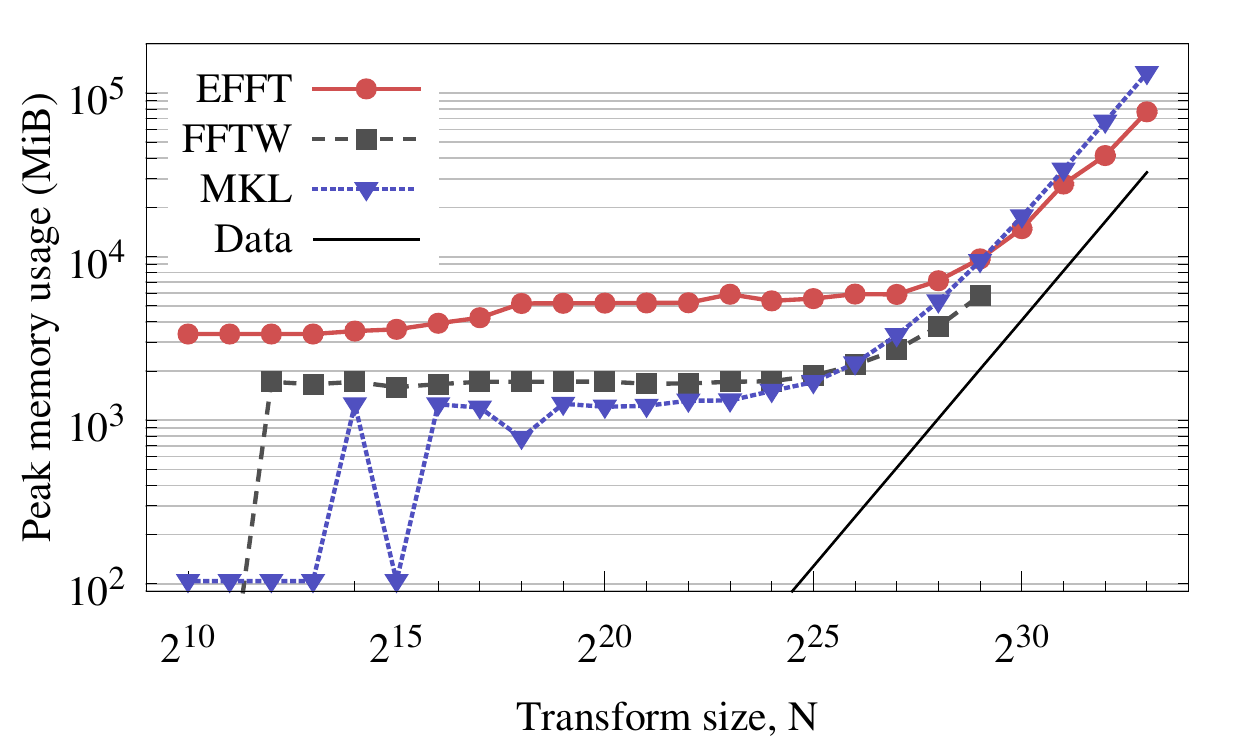}
\end{center}
\vskip -0.5em
\caption{Peak memory usage as a function of array size. We found that for most sizes the memory consumption of EFFT is comparable to memory consumption of MKL DFT.\label{fig:memory}}
\end{figure}

Figure~\ref{fig:memory} shows the peak virtual memory usage for each of the libraries. Memory usage was determined by querying the file \texttt{/proc/(pid)/status} and reading the line beginning with \texttt{VmPeak}. Here, \texttt{(pid)} is the process ID of the DFT implementation, and sampling rate was set at 1~kHz. For small array sizes, measurements were not consistent from run to run due to short run times. Solid black line in Figure~\ref{fig:memory} shows the amount of data in the input DFT array.

Figure (\ref{fig:memory}) shows that for small array sizes,  all libraries have memory usage around 2~GB. For larger arrays ($N>2^{28}$), EFFT uses less memory than MKL, however, both libraries require 3x to 4x as much memory as the input data array contains. FFTW has consistently lower memory consumption than EFFT for large arrays, around 2x the data size.

Finally, in order to ensure fair comparison of the three implementations, we estimate the accuracy of the transforms. This is done using the methodology proposed by the FFTW project \cite{fftw}. The data array is initialized with random values from $-0.5$ to $0.5$ and transformed. Then the computed result, $F$, is compared to the ``exact'' solution $F^e$ obtained using a single-threaded ``infinite precision'' calculation with the help of the GNU Multiple Precision (GMP) library. The comparison metric is the so-called L2 norm of the difference between the computed and ``exact'' DFT. Equation~(\ref{eq:l2norm}) defines the L2 norm.

\begin{equation}
\label{eq:l2norm}
\left|\left|F-F^e\right|\right|=\frac{\sqrt{\sum (F_k - F_k^e)^2}}{\sqrt{\sum{\left(F_k^e\right)^2}}}
\end{equation}

\begin{figure}[H]
\begin{center}
\includegraphics[width=\columnwidth]{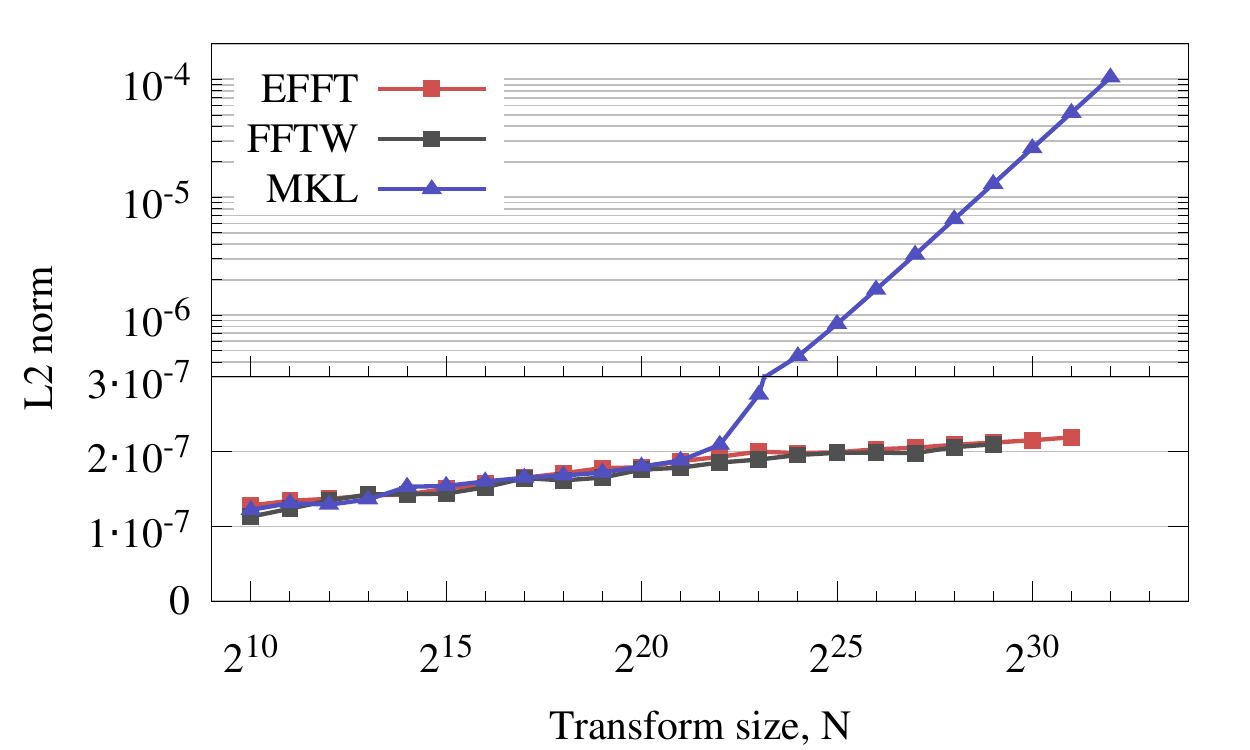}
\end{center}
\vskip -1em
\caption{Accuracy of DFTs  as a function of array size. \label{fig:l2norm}}
\end{figure}

Figure~\ref{fig:l2norm} reports the L2 norm of the deviation from the ``exact'' solution for power of 2 transform sizes with EFFT, MKL and FFTW.

All codes produce comparable L2 norms of order a few times $10^{-7}$, except for MKL for large transforms. Starting at the size $N=2^{22}$, the accuracy of MKL degrades and continues to degrade linearly towards greater sizes. Considering this aspect, EFFT built on small-size single-threaded MKL transforms with Intel Cilk Plus results in both better performance and better accuracy than the multi-threaded MKL DFT implementation.

\subsection{Parallel Scalability}\label{sec:Benchmarks-scalability}

While raw maximum performance reported in Section~\ref{sec:Benchmarks-performance} is important for practical application, it is also informative to study the parallel efficiency of the implementations. We define parallel efficiency, $\eta$, as the ratio of the actual performance with $T$ threads, $P(T)$, to the projected performance assuming linear speedup:
\begin{equation}
\eta(T) = \frac{P(T)}{T\times P(1)}.
\end{equation}
In compute-bound workloads, $\eta$ may remain constant up to values of $T$ equal to the number of physical cores in the processor.
However, for bandwidth-bound workloads like DFFT, $\eta$ is expected to decrease with increasing $T$. In order to evaluate the thread scalability of EFFT, we compare $\eta$ for EFFT with $\eta$ for the STREAM benchmark \cite{STREAM}.

\begin{figure}[H]
\begin{center}
\includegraphics[width=\columnwidth, clip=true, trim=0 0 0 1.5em]{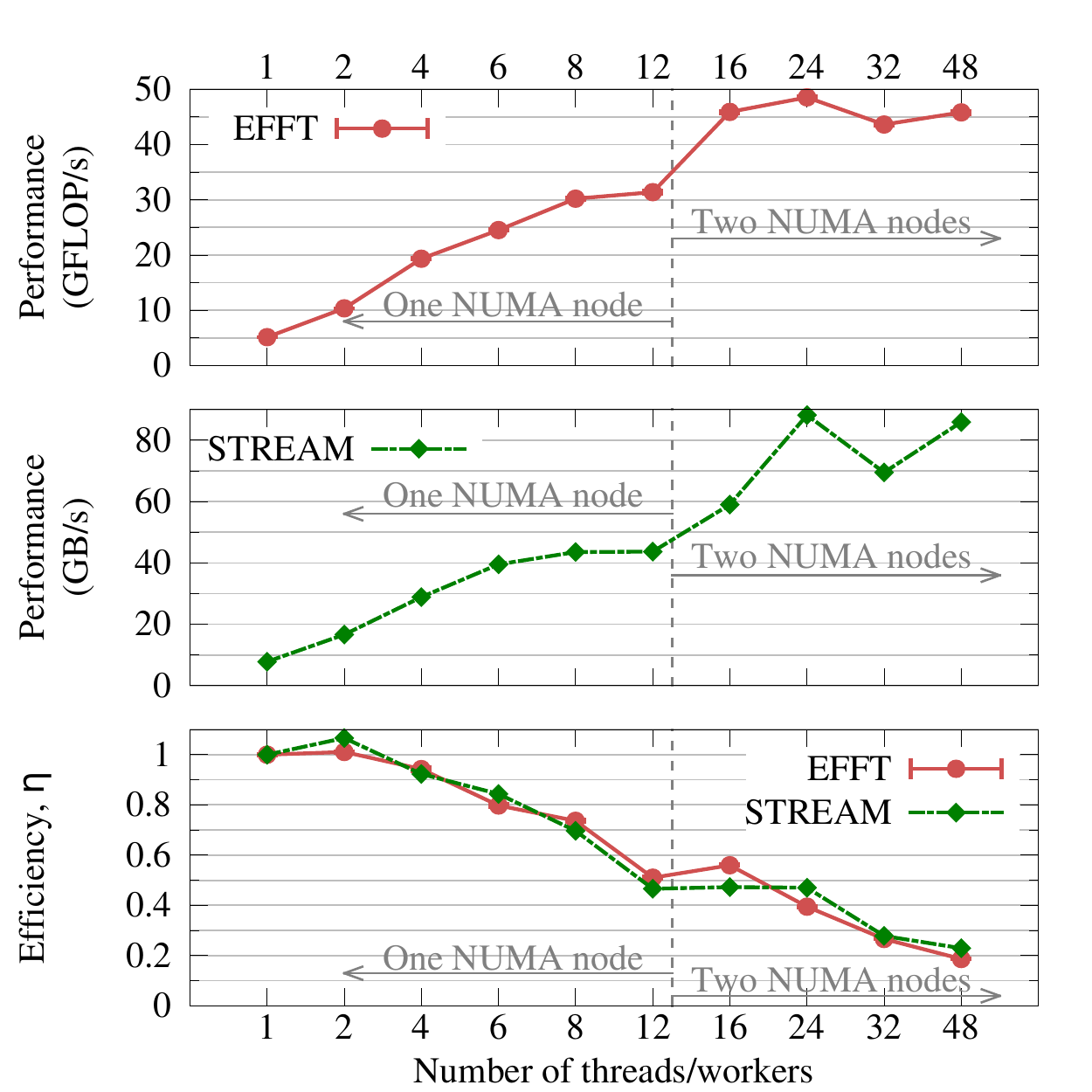}
\end{center}
\vskip -1em
\caption{Parallel scalability of EFFT and STREAM. \label{fig:parallel-perf}}
\end{figure}

Figure~\ref{fig:parallel-perf} reports the performance and parallel efficiency of EFFT and of the STREAM benchmark. For EFFT, $N=2^{30}$ was used. In all tests, thread affinity with $T$ threads was set to \texttt{KMP\_AFFINITY=explicit,proclist=[0-N]}, where $N=T-1$. This binds threads to individual processor cores in such a way that for $1<T\leq 12$, threads are placed on successive cores on CPU 0, and for $T>12$, threads start filling CPU 1. 
That said, for $T \leq 12$, only one NUMA node is used in our two-way system, and for $T>12$, two NUMA nodes are used.

According to Figure~\ref{fig:parallel-perf}, the efficiency of EFFT at the optimal value of $T=24$ threads is $\eta=0.39$, while STREAM has better efficiency, $\eta=0.47$. EFFT is expected to be less efficient than STREAM for $T=24$. That is because in STREAM, threads read and write only NUMA-local data (i.e., data mapped to memory banks local to the socket on which the thread is executing), while EFFT has a more complex pattern of memory traffic with non-local NUMA accesses. However, for $T=12$, when NUMA is not involved, EFFT has slightly better efficiency better than STREAM (which is consistent with data reuse in caches in EFFT). This observation indicates good parallel scalability of the bandwidth-bound EFFT algorithm.

\newpage
\section{Discussion}\label{sec:Conclusion}

Our goal in this work was to implement a multi-threaded code for very large ($N\gtrsim 10^9$) 1D DFTs. 
We achieved that goal by developing the EFFT library based on a serial MKL DFT implementation and Intel Cilk Plus to parallelize and vectorize the multi-threaded implementation. Our implementation performs better than multi-threaded MKL DFT for $N>2^{22}$ by 1.1x to 1.5x, with better accuracy of the calculation and lower memory usage than MKL. We also benchmarked EFFT against FFTW and measured 2x better performance with the former without any loss in accuracy; however, FFTW has lower memory footprint.

This paper also demonstrated multiple optimization techniques and discussed the reasoning behind the performance gain that each technique produced: 
\begin{enumerate}[-]
\tighterenum
\item improving temporal data locality via loop tiling,
\item improving spatial data locality (in-place algorithm),
\item strip-mining to vectorize transcendental math,
\item using array notation to vectorize stride-2 operations (``gather'' and ``scatter''),
\item using the \texttt{cilk\_spawn}/\texttt{cilk\_sync} extensions of Intel Cilk Plus to effect parallel recursion, and
\item using the \texttt{cilk\_for} extension for loop parallelism.
\end{enumerate}
The optimization methodology presented in this paper is applicable not only to parallel DFFTs, but to a wide variety of computational problems problems.

Futrhermore, this publication demonstrated the strength of Intel Cilk Plus for parallelizing a workload with a complex, multi-level pattern of parallelism.
While the OpenMP standard offers similar functionality (tasking and dynamic number of threads in parallel regions), in this application we were not able to achieve satisfactory performance results with OpenMP despite investing a greater development effort than we did with Intel Cilk Plus.
In contrast, development with Intel Cilk Plus required little programming effort and resulted in accelerated performance.

The good parallel scalability of EFFT (considering its bandwidth-bound nature) and its reliance on automatic vectorization and portable Intel Cilk Plus parallel framework promise high chances of adapting this application to the Intel MIC architecture. In a future publication, we will report on the possibility of accelerating what we call ``enormous Fourier transforms'' using Intel Xeon Phi coprocessors.

The product of the publication, the EFFT library, is available for free download \cite{efft}.

\optcolend

\end{document}